# Mathematical Analysis of Anthropogenic Signatures: The Great Deceleration

*By* Ron W. NIELSEN[†]

**Abstract**. Distributions of anthropogenic signatures (impacts and activities) are mathematically analysed. The aim is to understand the Anthropocene and to see whether anthropogenic signatures could be used to determine its beginning. A total of 23 signatures were analysed and results are presented in 31 diagrams. Some of these signatures contain undistinguishable natural components but most of them are of purely anthropogenic origin. Great care was taken to identify abrupt accelerations, which could be used to determine the beginning of the Anthropocene. Results of the analysis can be summarised in three conclusions. 1. Anthropogenic signatures cannot be used to determine the beginning of the Anthropocene. 2. There was no abrupt Great Acceleration around 1950 or around any other time. 3. Anthropogenic signatures are characterised by the Great Deceleration in the second half of the 20th century. The second half of the 20th century does not mark the beginning of the Anthropocene but most likely the beginning of the end of the strong anthropogenic impacts, maybe even the beginning of a transition to a sustainable future. The Anthropocene is a unique stage in human experience but it has no clearly marked beginning and it is probably not a new geological epoch.

## 1. Introduction

Steffen et al. (2004, pp. 132, 133) published an excellent sets of diagrams illustrating the rapidly increasing human activities in recent years. These diagrams became exceptionally popular but they have never been mathematically analysed. They contain a rich source of information, which has never been explored. By now, new data became available and my aim is to analyse them mathematically, because reliable and complete information can be obtained only by a rigorous examination of evidence.

Anthropogenic signatures (activities and impacts) have been rapidly increasing in recent years. Their primary driving force is the rapid growth of population but their combined intensity is reflected in the economic growth, which is closely linked with such anthropogenic signatures as the production and consumption of energy, consumption of fertilizers, pollution of the atmosphere, land and water, water consumption, land degradation, loss of tropical forests, consumption of marine resources, ocean acidification, stratospheric ozone depletion, increased transportation, increased consumption of renewable and non-renewable resources, interference with nitrogen and phosphorus cycles, and more. Maybe all anthropogenic impacts and activities are embodied in the economic growth but if not all, then many of them are. Economic growth appears to reflect the *combined* intensity of anthropogenic impacts and activities. It is hard, maybe even impossible, to decouple economic growth from environmental impacts. Recent study of

[†] Australian National University, Department of Nuclear Physics, Canberra, ACT, Australia (research in nuclear physics – retired); Griffith University, Environmental Futures Research Institute, Gold Coast, Queensland, Australia (research in economics and demography – resigned).

☎. +61407201175

✉. ronwnielsen@gmail.com

this issue resulted in the following conclusion: "It is therefore misleading to develop growth-oriented policy around the expectation that decoupling is possible" (Ward, Sutton, Werner, Costanza, Mohr, & Simmons, 2016, p. e0164733-10).

Anthropogenic impacts and activities are closely correlated with the growth of population. As observed by Waters, et al. (2016, p. aad2622-2) the "increase in the consumption of natural resources is closely linked with the growth of the human population." Correlations might not be linear but it is hardly surprising that with the rapidly increasing population human activities and impacts have been also rapidly increasing.

The rapid increase in the intensity of anthropogenic signatures in recent years is discussed extensively in my book (Nielsen, 2006). This apparently new phenomenon is described as the Anthropocene (see for instance Crutzen, 2004; Crutzen & Stoermer, 2000; Ehlers & Krafft, 2006; Steffen, Grinevald, Crutzen & McNeill, 2011; Zalasiewicz, Williams, Steffen & Crutzen 2010), which is now proposed to be recognised as a new geological epoch. However, all attempts to determine its beginning have been so far unsuccessful.

In order to accept the Anthropocene as a new geological epoch, a convincing stratigraphic evidence has to be presented but it would be also helpful to show that the beginning of the Anthropocene is convincingly demonstrated by a sharp acceleration in the intensity of anthropogenic signatures. They should have a clearly marked beginning.

Such a close examination of data has never been done before but according to Steffen et al. (2004), who compiled and published the sets of diagrams illustrating the rapidly increasing anthropogenic impacts and who examined them visually, there was a sharp increase in anthropogenic activities around 1950: "Sharp changes in the slope of the curves occur around the 1950s *in each case* and illustrate how the past 50 years have been a period of dramatic and unprecedented change in human experience" (Steffen, et al. 2004, p. 132; emphasis added). This description leaves no doubt about the interpretation of diagrams illustrating anthropogenic signatures. It is not a monotonic increase but an increase characterised by *a commonly occurring abrupt acceleration* around 1950.

A few years later, this postulate of sharp changes was reinforced by the introduction of the concept of the Great Acceleration (Hibbard, et al., 2007). According to these authors, "Since the 1950s, there has been a Great Acceleration in the scope, scale, and intensity of mutual impacts on the human–environment system" (Hibbard, et al., 2007, p. 343). Again, there is no doubt how the concept of the Great Acceleration is interpreted. It does not describe a gradual and rapid increase but a rapid *acceleration at a certain time*, which is around 1950.

It has been, therefore, natural to suggest that the Great Acceleration could be used to determine the beginning of the Anthropocene: "Although there has been much debate around the proposed start date for the Anthropocene, the *beginning* of the Great Acceleration has been a leading candidate (Zalasiewicz et al., 2012)" (Steffen, Broadgate, Deutsch, Gaffney & Ludwig, 2015, p. 83; emphasis added).

It is a reasonable suggestion but first it would have to be convincingly demonstrated that there *was* a great acceleration around 1950. This has never been done and no-one seems to be interested in doing it maybe because the concept of the Great Acceleration is by now so firmly established that no-one seems to question its validity. An excellent compilation of data describing time-dependent distributions of the intensity of individual anthropogenic impacts and activities was published by the International Geosphere-Biosphere Programme (Ludwig, 2014). It is interesting to notice that while great care was taken to present reliable data, the concept of the Great Acceleration is accepted without any reservation: "The second half of the 20th Century is unique in the history of human existence. Many human activities reached take-off points sometime in the 20th Century and *sharply accelerated* towards the end of the century. The last 60 years have *without doubt* seen the most profound transformation of the human relationship with the natural world in the history of humankind" (IGBP, 2015; emphasis added).

The beginning of the "last 60 years" is around 1950. Clearly, the article promotes the idea that "without doubt*"* "the most profound transformation of the human relationship with the natural world in the history of humankind" commended around 1950 and that this "profound transformation" is characterised by a *sharp acceleration* in human activities.

Concept of the Great Acceleration around 1950 was not only published recently in one of the most prestigious, peer-reviewed scientific journals (Waters et al., 2016) but also popularised by the Australian Broadcasting Corporation (ABC, 2016). The article published in *Science* (Waters, et al. 2016) in preparation for the final submission of the proposal for accepting the Anthropocene as the new geological epoch, contains a clear and unambiguous claim in its title: "The Anthropocene is functionally and stratigraphically distinct from the Holocene." In this article, concept of the Great Acceleration is accepted without any reservation and is repeatedly featured in various diagrams. Maybe it will be also featured in the final submission for the acceptance of the Anthropocene as a new geological epoch. "The inflection point at ~1950 CE coincides with the Great Acceleration (8, 9), a prominent rise in economic activity and resource consumption that accounts for the marked mid-20th century upturns in or inceptions of the anthropogenic signals" (Waters, et al., 2016, p. aad2622-3). Here again, there is no ambiguity about the way the Great Acceleration is understood and described. It is an acceleration at a specific time and this specific time is around 1950.

The "prominent rise in economic activity and resource consumption" around 1950 would have to be convincingly demonstrated before claiming it in support of the Anthropocene as a new geological epoch.

It is my aim now to analyse mathematically distributions describing recent anthropogenic signatures in order to understand their dynamics. In my analysis I shall use the excellent compilation of data prepared and published by Ludwig (2014), with the exception of the data describing the growth of population and economic growth, because in the past few years I have carried out extensive investigation of these issues (Nielsen, 2014, 2015, 2016a, 2016b, 2016c, 2016d, 2016e, 2016f, 2016g, 2016h, 2016i, 2016j, 2016k, 2016l, 2016m, 2016n, 2016o, 2016p, 2017a, 2017b, 2017c, 2017d, 2017e, 2017f, 2017g). All this research was essential for the correct understanding of the Anthropocene.

This surge of publications in the last two years represents an accumulation of my earlier papers, which I could not publish because Griffith University does not give financial support for the publication of research results. Papers may be accepted for publication but they are not published until authors can find a free outlet, which is not always easy.

My analysis of the growth of population and of economic growth was supported by a full set of data presented in the following publications: Biraben, 1980; Birdsell, 1972; Clark,1968; Cook,1960; Deevey, 1960; Durand, 1974; Gallant, 1990; Hassan, 2002; Haub, 1995; Livi-Bacci, 1997; Maddison, 2010; McEvedy & Jones, 1978; Taeuber & Taeuber, 1949; Thomlinson, 1975; Trager, 1994, United Nations, 1973, 1999, 2013; US Census Bureau, 2017. Some of the historical data were conveniently compiled by Manning (2008) and by US Census Bureau (2017).

Analysis of the remaining anthropogenic signatures is based on the data published by the International Geosphere-Biosphere Programme (Ludwig, 2014). References to the respective individual sets of data are listed in her compilation. I have also included data for the carbon dioxide emissions from fossil fuels (EPI, 2013).

The primary aim of my study is to understand the *past* growth and to identify sudden accelerations, which could be used *in support* of the concept of the Great Acceleration around 1950. My aim is not to predict growth. Such investigations would have to be carried out separately. Possible future trajectories will be presented but only to understand the overall dynamics of growth.

## 2. Definitions

Growth of population and economic growth ware increasing hyperbolically in the past 2,000,000 years (Nielsen, 2017c). Hyperbolic growth is described by the reciprocal of a decreasing linear function:

$$S(t) = \frac{1}{a - kt}. \tag{1}$$

where $S(t)$ is the size of the growing entity, $a$ and $k$ are positive constants and $t$ is the time.

For a good quality data over a sufficiently large range, hyperbolic growth can be uniquely identified by the decreasing liner distribution of its reciprocal values (Nielsen, 2014, 2017f):

$$\frac{1}{S(t)} = a - kt, \tag{2}$$

Hyperbolic distributions escape to infinity at a fixed time when $t = a/k$ or when their reciprocal values are zero.

Extended or higher-order hyperbolic growth can be described by the reciprocal of higher order polynomials.

$$S(t) = \frac{1}{\sum_{i=0}^{n>1} a_i t^i}. \tag{3}$$

If restriction for $n$ is removed, then eqn (3) includes also the first-order hyperbolic distribution described by the eqn (1).

Exponential growth is described by the following equation:

$$S(t) = c e^{rt}, \tag{4}$$

where $c$ is the normalisation constant related to the constant of integration and $r$ is the growth rate.

For a good quality data over a sufficiently large range, exponential growth can be uniquely identified by the straight line in the semilogarithmic display because

$$\ln S(t) = \ln c + rt. \tag{5}$$

Extended or higher-order exponential growth is described by the following equation:

$$S(t) = \exp\left[\sum_{i=0}^{n>1} a_i t^i\right]. \tag{6}$$

The normalisation constant is given by $a_0$. Equation (6) can be described as the higher order exponential distribution depending on the order of the polynomial.

Growth rate $R$ is defined as

$$R = \frac{1}{S(t)} \frac{dS(t)}{dt}. \tag{7}$$

For a discrete set of values, it is calculated using the following formula:

$$R_{i+1} = \frac{1}{S_i} \frac{S_{i+1} - S_i}{t_{i+1} - t_i}. \tag{8}$$

Growth rates can be used to find appropriate mathematical descriptions of data (Nielsen, 2017e).

Growth rate can be represented as a function of *time* or as a function of the *size* of the growing entity. If it is represented as a function of time, then there is a general solution to the relevant differential equation, which can be used to calculate mathematical distribution describing growth.

If

$$\frac{1}{S(t)} \frac{dS(t)}{dt} = f(t), \tag{9}$$

then

$$S(t) = \exp \int f(t) dt. \tag{10}$$

If growth rate is represented as

$$f(t) = a + bt, \tag{11}$$

then

$$S(t) = \exp(at + 0.5bt^2 + C), \tag{12}$$

where $C$ is a constant of integration, which is determined by comparing calculated curve with data.

This expression can be presented as

$$S(t) = \exp(a_0 + a_1 t + a_2 t^2). \tag{13}$$

It is a second-order exponential distribution. If $a_2 > 0$, i.e. if growth rate is described by an increasing linear function, then the distribution described by eqn (13) continues to increase indefinitely with time. In this case, the second-order exponential distribution is continually *accelerating*. If $a_2 < 0$, then the distribution described by eqn (13) will reach a maximum and will start to decrease. It is a distribution, which is continually *decelerating*.

Parameters $a_1$ and $a_2$ determine the shape of the distribution, while parameter $a_0$ is just the normalization constant, which has to be determined by comparing the calculated distribution with data.

If growth rate decreases exponentially with time, i.e. if

$$\frac{1}{S(t)}\frac{dS(t)}{dt} = ae^{bt}, \qquad (14)$$

then (Nielsen, 2017e)

$$S(t) = C\exp\left[\frac{a}{b}e^{bt}\right]. \qquad (15)$$

It is a pseudo-logistic growth because, with the increasing time, the size the growing entity increases asymptotically to the constant $C$. This is one of the two types of trajectories, which describe the current growth of the world population (Nielsen, 2017e).

If growth rate depends linearly on the *size* of the growing entity, i.e. if

$$\frac{1}{S}\frac{dS}{dt} = a_0 + a_1 S, \qquad (16)$$

and if $a_0 \neq 0$, then (Nielsen, 2017e)

$$S(t) = \left[Ce^{-a_0 t} - \frac{a_1}{a_0}\right]^{-1}, \qquad (17)$$

where

$$C = \left[\frac{1}{S_0} + \frac{a_1}{a_0}\right]e^{a_0 t_0}. \qquad (18)$$

If $a_0 > 0$ and $a_1 < 0$, i.e. if growth rate is decreasing linearly with the size of the growing entity, then eqn (17) describes logistic growth.

If $a_1 > 0$, then eqn (17) describes a pseudo-hyperbolic growth, which escapes to infinity at a fixed time.

If $a_0 = 0$, then eqn (16) is

$$\frac{1}{S}\frac{dS}{dt} = a_1 S \qquad (19)$$

It describes the first-order hyperbolic growth and has to be solved separately. Its solution is by substitution $Z = S^{-1}$. If eqn (19) is expressed as

$$\frac{1}{S}\frac{dS}{dt} = kS, \qquad (20)$$

then its solution is given by eqn (1).

If $k$ is not constant but is assumed to depend on time, i.e. if $k$ is replaced by $k(t)$, then the eqn (20) is

$$\frac{1}{S}\frac{dS}{dt} = k(t)S, \qquad (21)$$

and its solution is

$$S(t) = -\left[\int k(t)dt\right]^{-1}. \qquad (22)$$

If $k(t)$ is represented by a polynomial, then the eqn (22) can be expressed as eqn (3).

I shall always use the simplest mathematical descriptions of growth trajectories, as described in Table 1.

**Table 1**. *The simplest representations of data used in the analysis of anthropogenic signatures.*

| Growth Rate | Growth | Growth Trajectory |
|---|---|---|
| $R(t) = r$ | Exponential | $S(t) = ce^{rt}$ |
| $R(t) = a_1 + a_2 t$ | Second-order exponential | $S(t) = \exp(a_0 + a_1 t + a_2 t^2)$ |
| $R(S) = kS$ | Hyperbolic | $S(t) = (a - kt)^{-1}$ |
| $R(S) = a_0 + a_1 S$, $a_0 \neq 0$, $a_1 > 0$ | Pseudo-hyperbolic | $S(t) = \left[Ce^{-a_0 t} - \dfrac{a_1}{a_0}\right]^{-1}$ |
| $R(S) = a_0 + a_1 S$, $a_0 > 0$, $a_1 < 0$ | Logistic | $S(t) = \left[Ce^{-a_0 t} - \dfrac{a_1}{a_0}\right]^{-1}$ |
| $R(t) = ae^{bt}$ | Pseudo-logistic | $S(t) = C\exp\left[\dfrac{a}{b}e^{bt}\right]$ |

$R(t)$ – growth rate expressed as a function of time; $R(S)$ – growth rate expressed as a function of the size of the growing entity

Mathematical formulae listed in Table 1 are derived by using the simplest representations of growth rates, which are in general linear. However, I have also included an exponential representation, which is also relatively simple and applies, for instance, to the recent growth of the world population (Nielsen, 2017e).

It should be also noted that the general trend of distributions describing growth is determined only by the general trend of growth rates (Nielsen, 2017e). In general, fluctuation and oscillations in growth rates have no impact on the general trends of growth trajectories. This relation between growth rates and the corresponding distributions describing growth simplifies analysis of data.

If good quality data are available, growth rate should be always calculated and analysed because it can assist in a unique identification of growth trajectories. For instance, data might be represented equally well by a linear distribution, exponential distribution or by a second-order exponential distribution. In such a case, the usual procedure should be to select the linear distribution because it is the simplest representation of data. However, if growth rate *increases* with time, then the only option, out of these three possibilities, is the more complicated second-order exponential distribution because for the linear distribution the growth rate *decreases* hyperbolically with time and for the exponential distribution it is *constant*.

It is also important to understand that the decreasing growth rate does not necessarily describe the decreasing growth trajectories. As long as growth rate is on average positive, the corresponding growth trajectory will be increasing. If growth rate is on average constant or increasing, the corresponding growth will follow an increasing and *accelerating* trajectory. If growth rate is on average positive and decreasing, growth trajectory will be also increasing but *decelerating*. Growth trajectory will be decreasing only if growth rate is on average negative.

## 3. Analysis of data

### *3.1. Growth of population and economic growth*

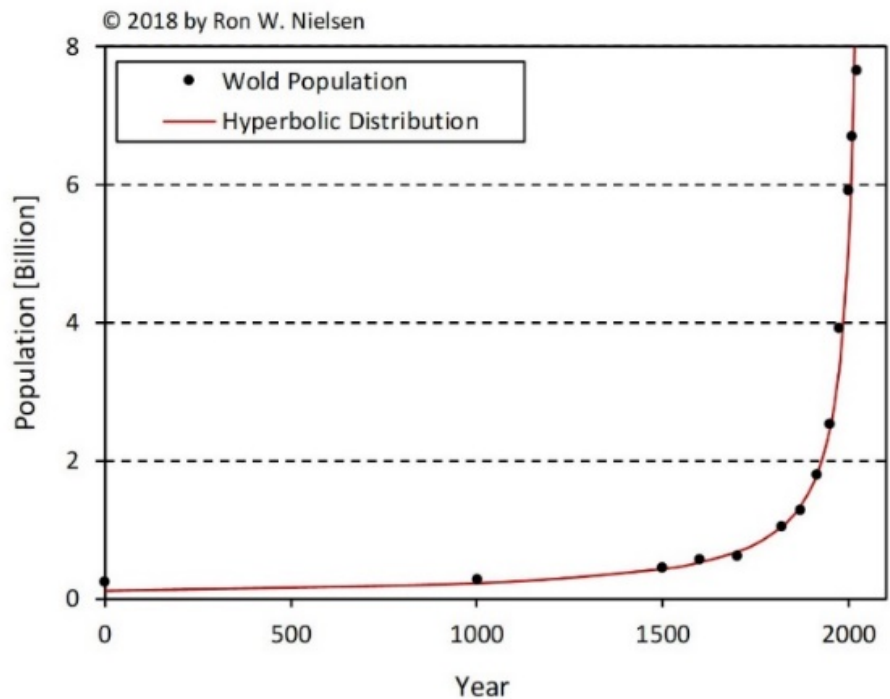

**Figure 1.** *Data describing growth of the world population (Maddison, 2010) follow closely a hyperbolic distribution defined by parameters* $a = 8.724 \times 10^0$ *and* $k = 4.267 \times 10^{-3}$. *Growth of population was not exponential, as first interpreted by Malthus (1798) but hyperbolic. There was no population explosion but a* monotonic *transition from a slow to fast growth. The perceived population explosion was just the natural continuation of hyperbolic growth. The fast growth of population has no mathematically determinable beginning.*

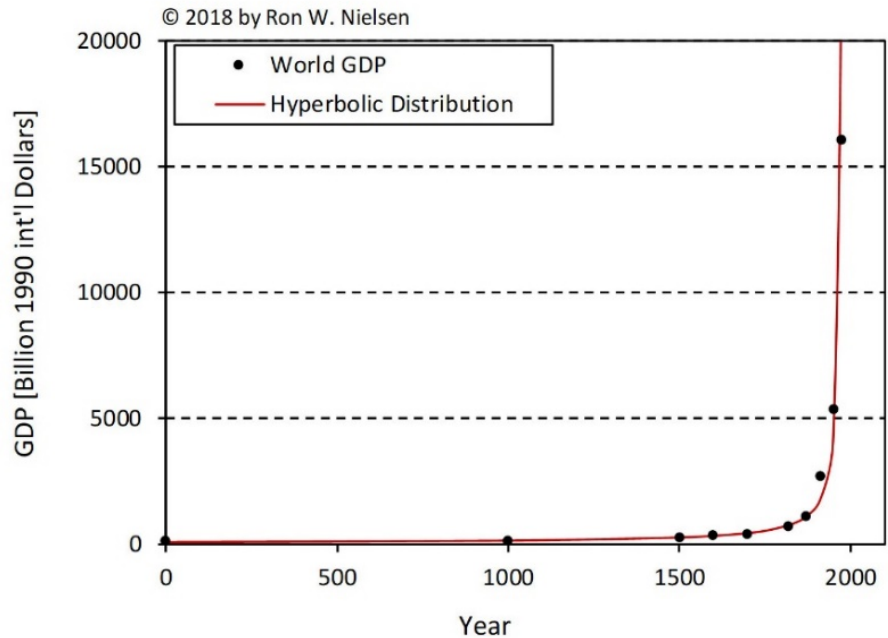

**Figure 2.** *Data describing growth of the world Gross Domestic Product (GDP) (Maddison, 2010), expressed in billions of 1990 international Geary-Khamis dollars, are compared with hyperbolic distribution defined by parameters $a = 1.716 \times 10^{-2}$ and $k = 8.671 \times 10^{-6}$ Growth of the world GDP was increasing monotonically along a hyperbolic trajectory. The fast growth in recent years had no mathematically determinable beginning suggesting strongly that the fast-increasing combined effects of anthropogenic activities, which are reflected in the economic growth, also had no mathematically determinable beginning.*

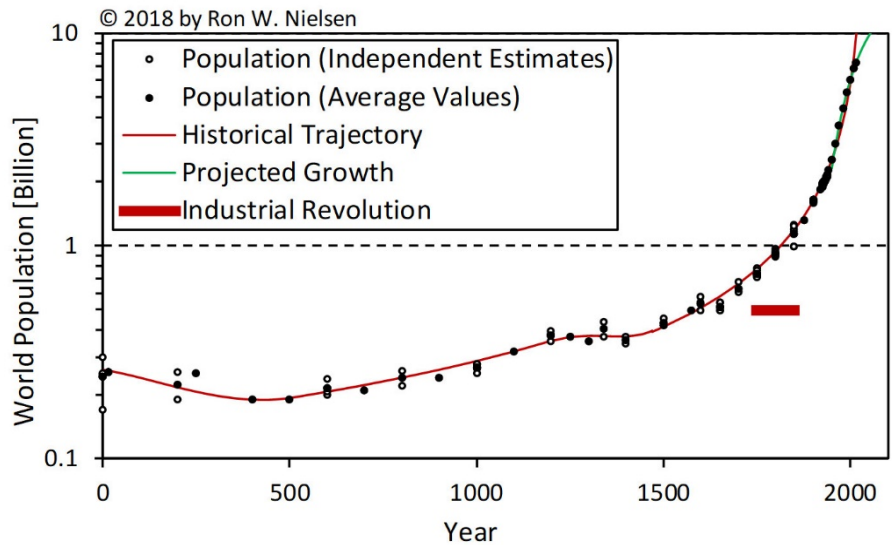

**Figure 3.** *Growth of the world population (Nielsen, 2016c, 2017c) during the AD time. Data from a complete set are used (see text). Calculated trajectory accounts for the major transition between two hyperbolic trajectories (425 BC – AD 510) and for a minor disturbance of the hyperbolic growth, which occurred between AD 1195 and AD 1470. Industrial Revolution had no impact on shaping growth trajectory. Parameters describing these calculations are listed in my earlier publications (Nielsen, 2016c, 2017c). There was no Great Acceleration around 1950.*

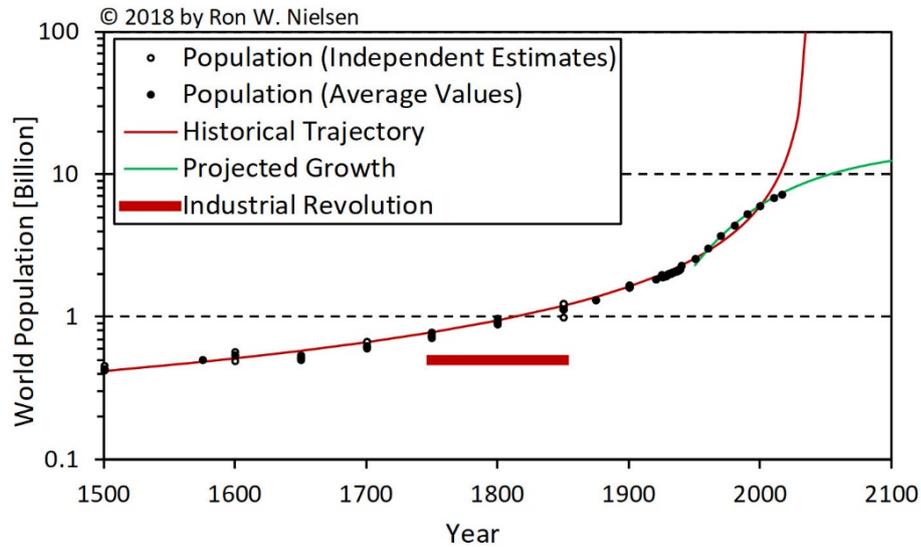

**Figure 4.** *Magnified section of the growth of population presented in Figure 3. There was no Great Acceleration around 1950 or at any other time but only a minor and temporary boosting. From around 1963, growth of the world population follows a continually decelerating trajectory (Nielsen, 2017e; United Nations, 2015; US Census Bureau, 2017).*

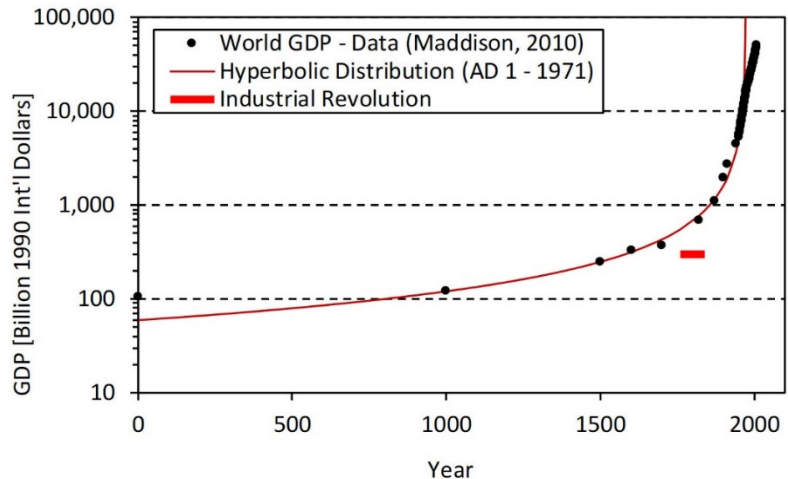

**Figure 5.** *Data for the world Gross Domestic Product (GDP) between AD 1 and 2008 (Maddison, 2010), expressed in billions of 1990 international Geary-Khamis dollars, are compared with hyperbolic distribution. The recent fast-increasing economic growth had no mathematically determinable beginning, strongly suggesting that the recent fast increasing intensity of the combined effects of anthropogenic activities, which are reflected in the economic growth, also had no mathematically determinable beginning. Industrial Revolution had no impact on shaping economic growth trajectory, even in Western Europe and even in the United Kingdom (Nielsen, 2014, 2016b, 2016l, 2017f). There was no Great Acceleration around 1950. On the contrary, from around 1950, growth of the world GDP, which represents total consumption of natural resources, started to be diverted to a slower trajectory, which was initially decelerating but is now following an accelerating exponential distribution (Nielsen, 2015). The new exponential trajectory is slower than the historical hyperbolic distribution.*

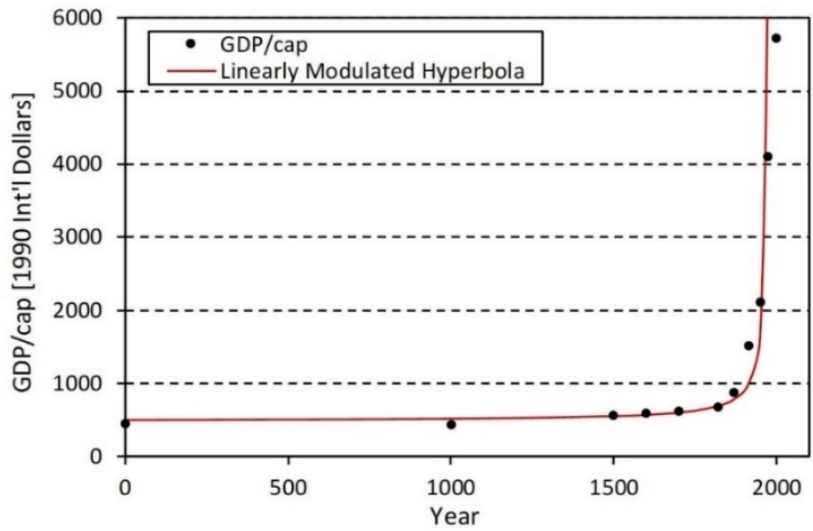

**Figure 6.** *Distribution describing income per capita (GDP/cap) obtained by dividing distributions shown in Figures 1 and 2. The best fit to the data is represented by a* monotonically *increasing linearly-modulated hyperbolic distribution (Nielsen, 2017a). Income per capita was approximately constant in the past but most recently it was fast increasing, However, there was no sudden transition from the slow to fast growth. These results suggest that the* combined *effects anthropogenic activities* per person *might have been also approximately constant in the past but they were gradually increasing. There is no mathematically determinable beginning of their fast increase.*

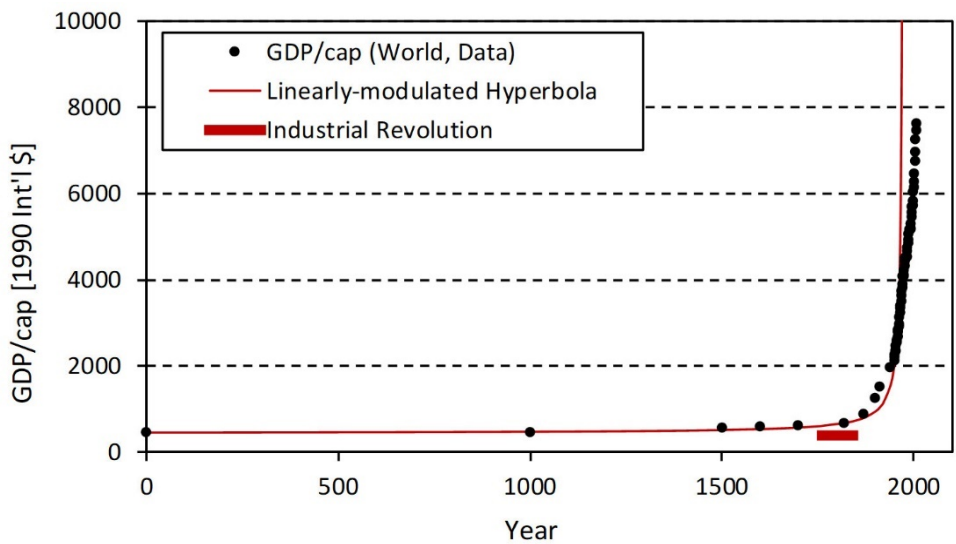

**Figure 7.** *Extended data describing growth of income per capita (GDP/cap) (Maddison, 2010) are compared with the linearly-modulated hyperbolic distribution (Nielsen, 2017a). There was no Great Acceleration in 1950 but a* deceleration *and diversion to a* slower *trajectory.*

## 3.2. Foreign direct investment

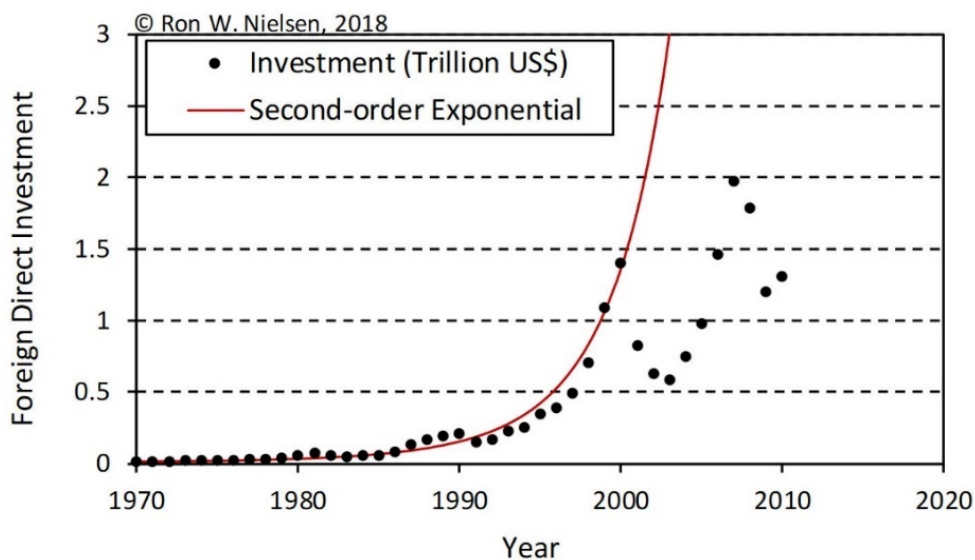

**Figure 8.** *Global foreign direct investment (FDI) in trillions of the current US$. The simplest mathematical representation of growth is by the second-order exponential distribution based on the simplest liner representation of the growth rate. Its parameters are: $a_0 = 1.380 \times 10^4$, $a_1 = -1.405 \times 10^1$ and $a_2 = 3.575 \times 10^{-3}$. It was a gradually accelerating trajectory because $a_2 > 0$. Growth was oscillating around a monotonically increasing trajectory until 2000. From that year, it became strongly unstable and unpredictable. There was no abrupt acceleration at any time but there are signs of deceleration and of a diversion to a slower pattern of growth.*

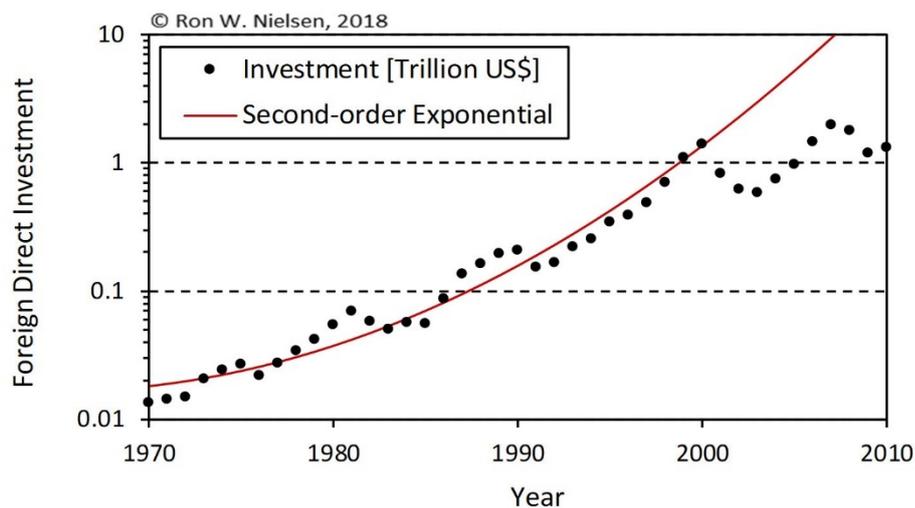

**Figure 9.** *Global foreign direct investment displayed using semilogarithmic scales of reference, showing more clearly that growth was oscillating around a monotonically increasing trajectory and that from around 2000 it started to follow a generally slower pattern of growth.*

## 3.3. Growth of urban population

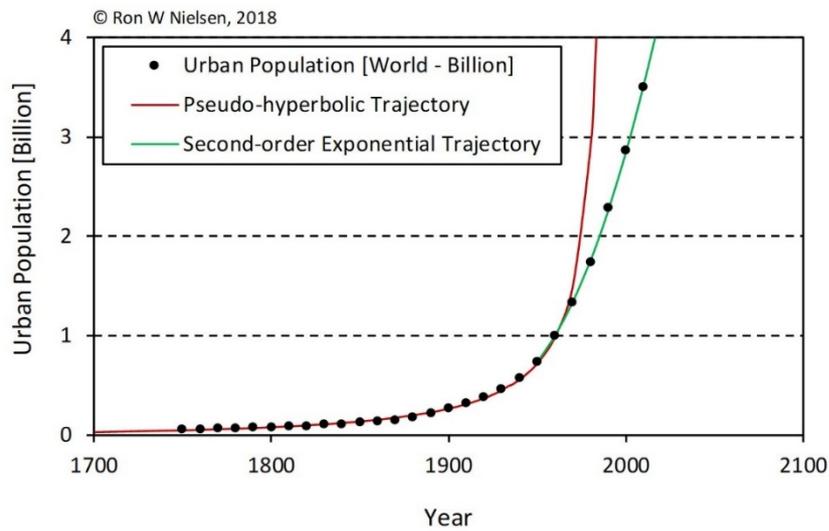

**Figure 10.** *Growth of the world urban population in billions. Initially, and up to around 1960, growth rate was increasing linearly with the size of human population. Growth of urban population was following a pseudo-hyperbolic trajectory defined by parameters $C = 1.536 \times 10^7$, $a_1 = 7.658 \times 10^{-3}$ and $a_2 = 2.798 \times 10^{-2}$. From around 1960 it was diverted to a slower and gradually decelerating second-order exponential trajectory as indicated by the linearly decreasing growth rate. Parameters of this slower trajectory are: $a_0 = -4.345 \times 10^2$ $a_1 = 4.133 \times 10^{-1}$ and $a_2 = -9.776 \times 10^{-5}$. There was no Great Acceleration around 1950 but a deceleration around 1960 and a commencement of a gradually decelerating trajectory.*

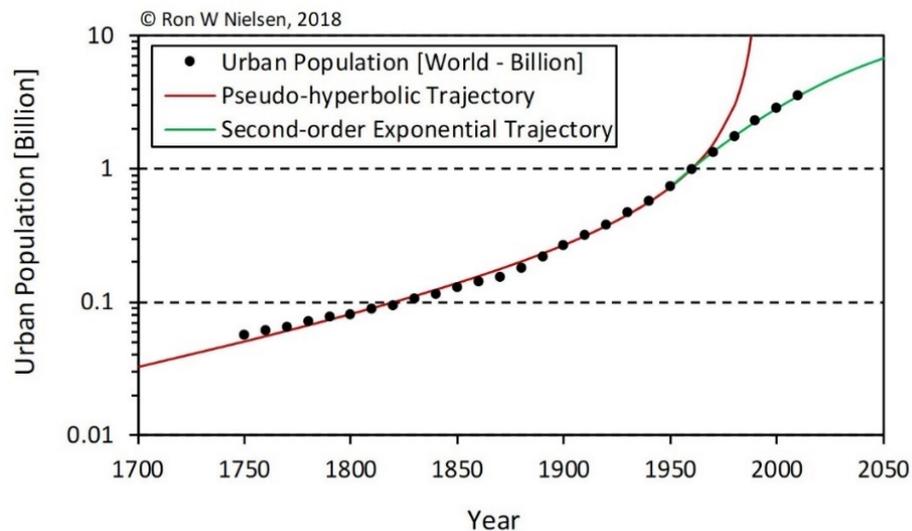

**Figure 11.** *Semilogarithmic display of the growth of the global urban population. This graph shows more clearly that the growth of urban population was following a monotonically increasing trajectory and that it was diverted to a slower and gradually decelerating trajectory. There was no Great Acceleration around 1950 or around any other time. On the contrary there was an abrupt deceleration around 1960 and a diversion to a slower, continually decelerating, trajectory.*

## 3.4. Consumption of primary energy

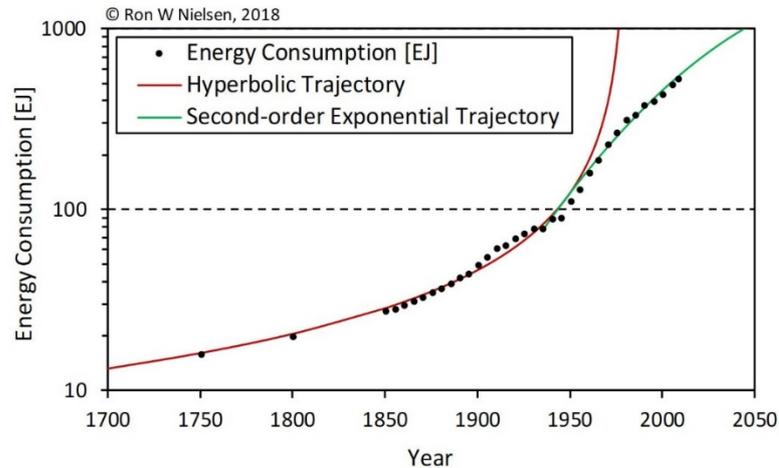

**Figure 12.** *Global consumption of primary energy in exajoules [EJ]. Consumption was increasing hyperbolically until 1950. This trajectory was indicated by the analysis of reciprocal values of data and of the growth rate. Its parameters are: $a = 5.356 \times 10^{-1}$ and $k = 2.705 \times 10^{-4}$. Around 1950, global consumption of primary energy was* decelerated *and diverted to a gradually* decelerating *second-order exponential trajectory. It parameters are: $a_0 = -3.955 \times 10^2$, $a_1 = 3.799 \times 10^{-1}$ and $a_2 = -8.954 \times 10^{-5}$. There was no Great Acceleration around 1950 but a deceleration and a diversion to a slower, continually decelerating trajectory.*

## 3.5. Consumption of fertilizers

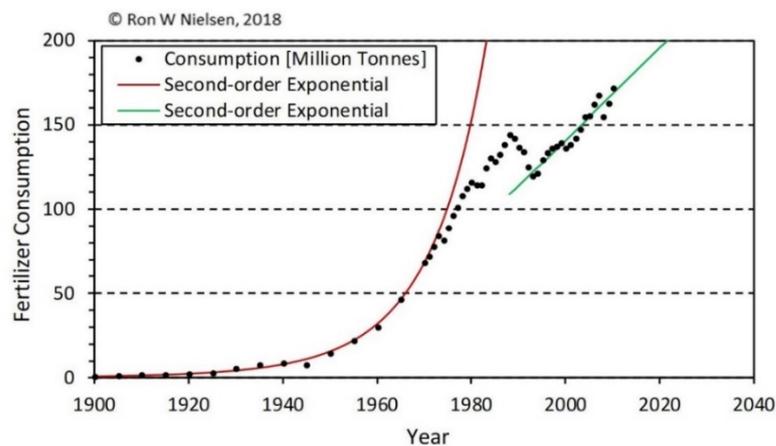

**Figure 13.** *Global consumption of fertilizers (in million tonnes) was increasing monotonically by following a second-order exponential trajectory ($a_0 = 7.788 \times 10^2$, $a_1 = -8.646 \times 10^{-1}$ and $a_2 = 2.393 \times 10^{-4}$) until around 1972 when it was* decelerated *and diverted to a* slower *trajectory. The slower growth reached an unexpected maximum around 1989 and the consumption of fertilizers started to decrease. The decline continued until around 1993 when the consumption of fertilizers started to follow an even* slower *and gradually* decelerating *second-order exponential trajectory described by parameters $a_0 = -6.341 \times 10^2$, $a_1 = 6.195 \times 10^{-1}$ and $a_2 = -1.500 \times 10^{-4}$. The was no Great Acceleration around 1950 or around any other time.*

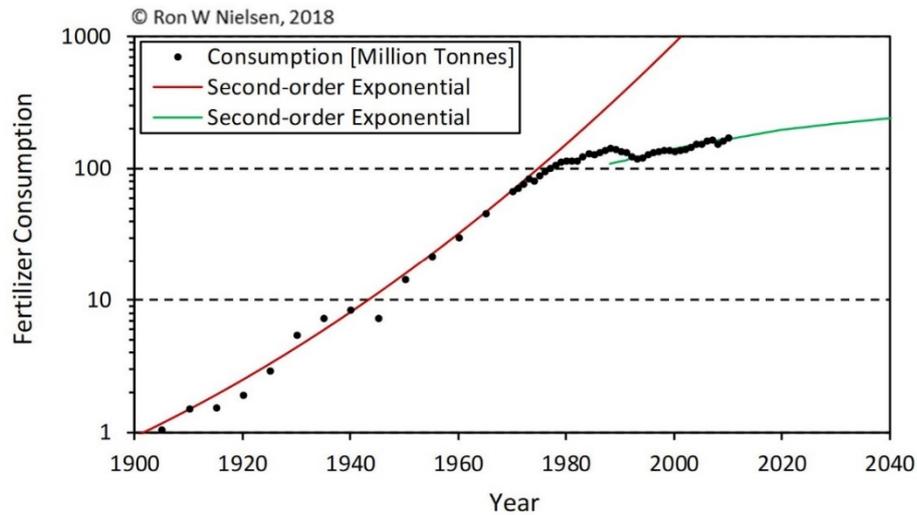

**Figure 14.** *Global consumption of fertilizers (in million tonnes) is shown here using semilogarithmic scales of reference. This diagram shows clearly that there was no abrupt acceleration in the growth trajectory around 1950. It also demonstrates that there was a* deceleration *around 1972 and a diversion to a* slower *trajectory, which reached an unexpected maximum, started to decrease and increase again along an even* slower *and gradually* decelerating *trajectory.*

## 3.6. Large dams

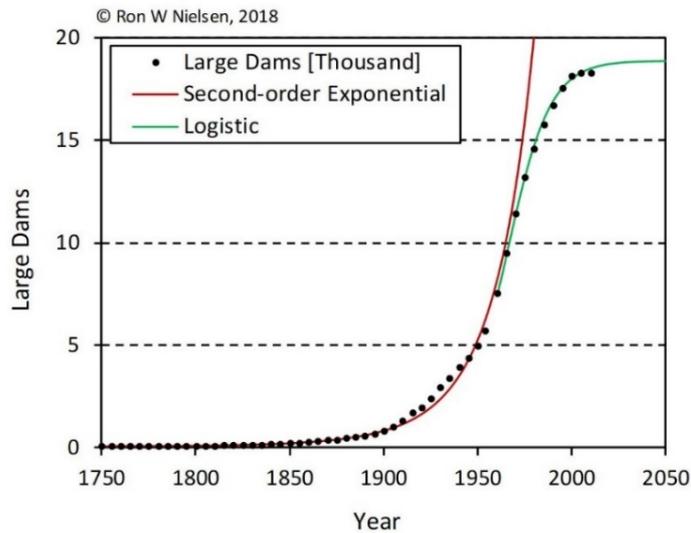

**Figure 15.** *Global number of existing large dams, in thousands. Growth rate was initially increasing along a linear trajectory, indicating a monotonically increasing and gradually accelerating second-order exponential growth of the number of existing large dams ($a_0 = 2.734 \times 10^2$, $a_1 = -3.208 \times 10^{-1}$ and $a_2 = 9.305 \times 10^{-5}$). From around 1965, growth rate was decreasing linearly with the number of existing dams, indicating logistic growth defined by parameters: $C = 7.642 \times 10^{72}$, $a_0 = 8.689 \times 10^{-2}$ and $a_1 = -4.599 \times 10^{-3}$. The calculated limit to growth is 18.9 thousand. The value recorded for 2010 is 18.3. There was no Great Acceleration in 1950 or at any other time.*

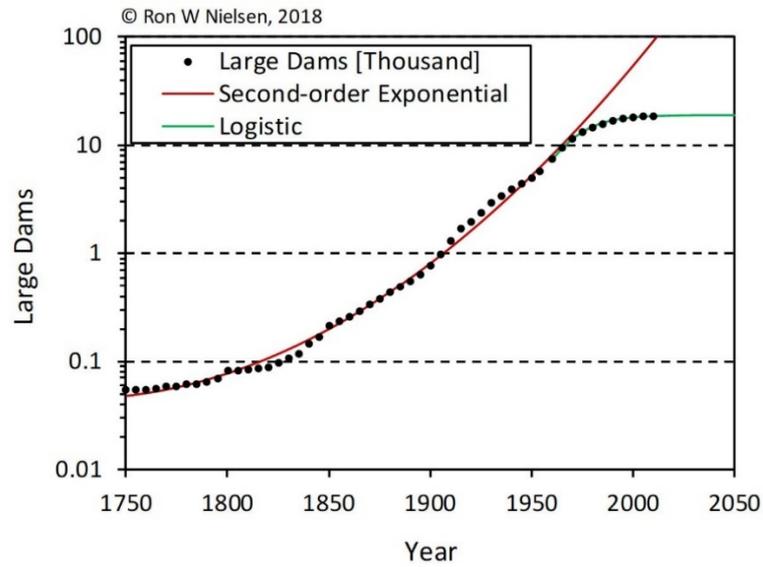

**Figure 16.** *Global number of existing large dams in thousands displayed here by using semilogarithmic scales of reference. Growth was increasing monotonically until around 1965 when it was diverted to a* slower *and gradually* decelerating *trajectory. There was no Great Acceleration in 1950 or at any other time.*

## 3.7. Water consumption

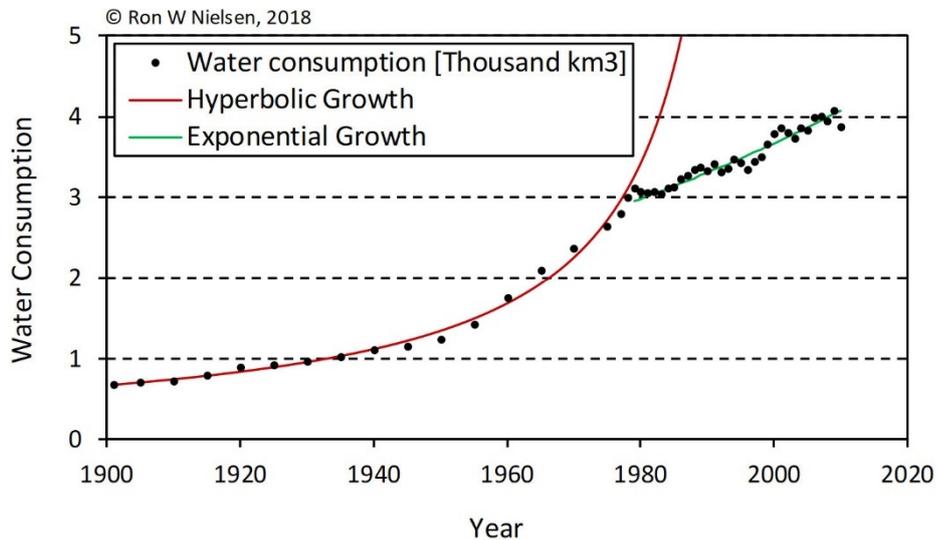

**Figure 17.** *Global consumption of water, in cubic kilometres, was increasing hyperbolically until around 1979. Its parameters are: $a = 3.025 \times 10^1$ and $k = 1.513 \times 10^{-2}$. In around 1979, world water consumption was* decelerated. *From that time on, growth rate was small and constant generating a slow exponential growth ( $c = 3.310 \times 10^{-9}$, $r = 1.041 \times 10^{-2}$ ). It is a slowly* accelerating *growth. Growth rate is only 1.04% per year. There was no Great Acceleration in 1950 or at any other time.*

## 3.8. Paper production

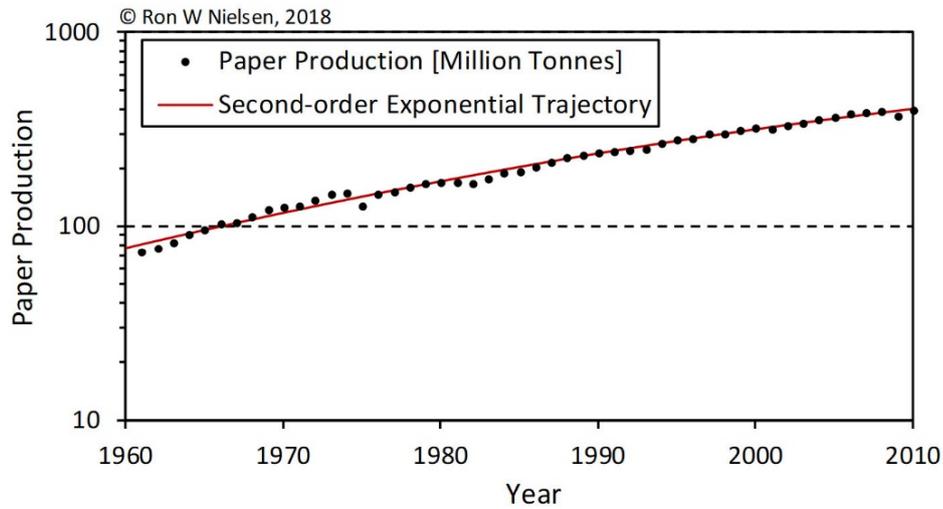

**Figure 18.** *Paper production was following a gradually* decelerating*, second-order, exponential trajectory* ( $a_0 = -9.279 \times 10^2$, $a_1 = 9.072 \times 10^{-1}$, $a_2 = -2.202 \times 10^{-4}$ ). *There was no abrupt acceleration at any time.*

## 3.9. Transportation

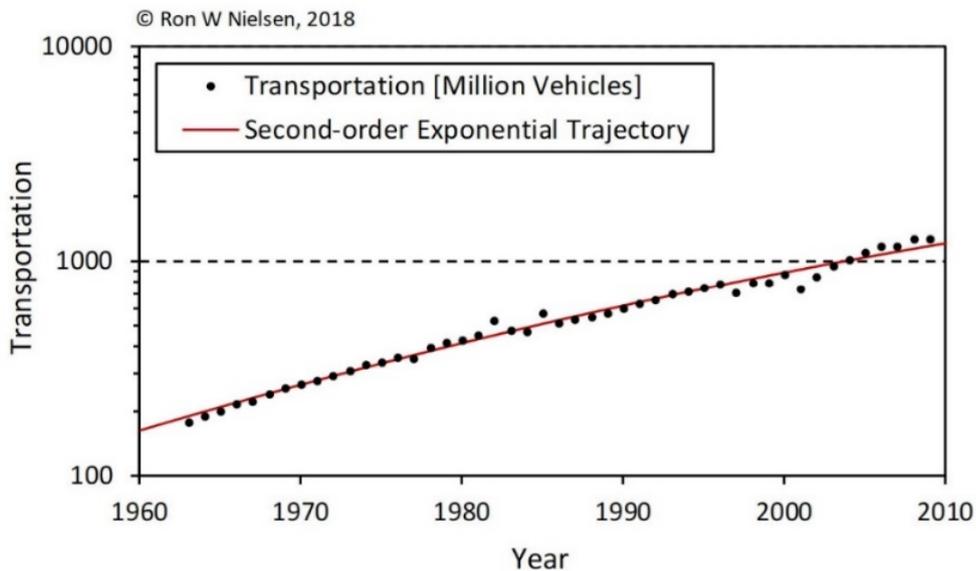

**Figure 19.** *Global transportation has been increasing along a gradually* decelerating *second-order exponential trajectory described by the following parameters:* $a_0 = -9.279 \times 10^2$, $a_1 = 9.072 \times 10^{-1}$ *and* $a_2 = -2.202 \times 10^{-4}$.

*3.10. Telecommunication*

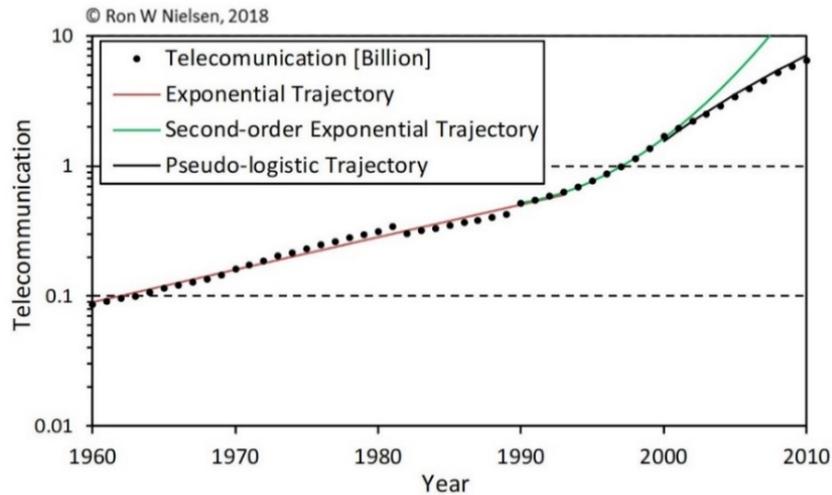

**Figure 20.** *Global telecommunication (billions of landlines and subscriptions). The three components of growth can be identified only by the analysis of the growth rate. Telecommunication was increasing exponentially until 1991 ( $c = 3.944 \times 10^{-51}$, $r = 5.798 \times 10^{-2}$ ). Around 1991, growth rate increased abruptly and growth was* accelerating *along a faster, second-order, exponential trajectory ( $a_0 = 2.945 \times 10^4$, $a_1 = -2.964 \times 10^1$, $a_2 = 7.457 \times 10^{-3}$ ). From around 2000, growth rate was decreasing and could be approximated either by a linear or exponential trajectory. Exponential trajectory was chosen because it is more likely that the number of landlines and subscription will be approaching asymptotically a certain constant value rather than that it will be irrevocably decreasing. The current trajectory is tentatively described by pseudo-logistic distribution ( $C = 2.200 \times 10^2$, $a = 8.256 \times 10^{30}$, $b = -3.645 \times 10^{-2}$ ). There was no Great Acceleration around 1950 but only a temporary acceleration commencing around 1991 and ending around 2000. This small but sustained acceleration is revealed only by studying the growth rate.*

*3.11. International tourism*

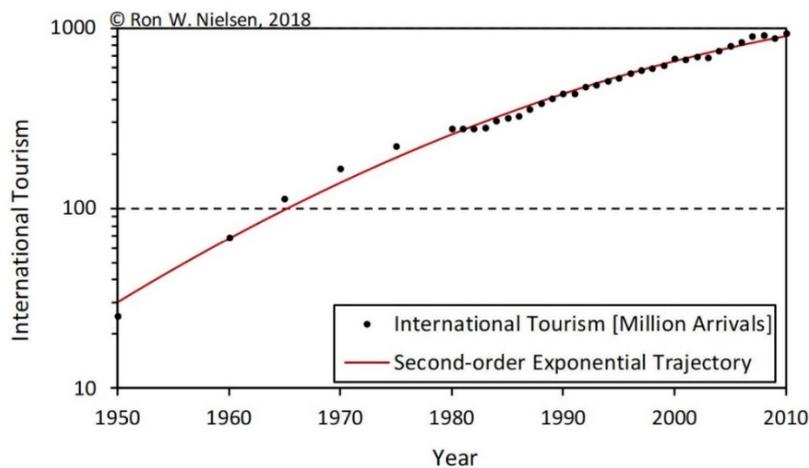

**Figure 21.** *Global international tourism was increasing along a* decelerating *second-order exponential trajectory ( $a_0 = 2.060 \times 10^3$, $a_1 = 2.030 \times 10^0$, $a_2 = -4.982 \times 10^{-4}$ ). There was no abrupt acceleration at any time.*

## 3.12. Atmospheric concentration of carbon dioxide

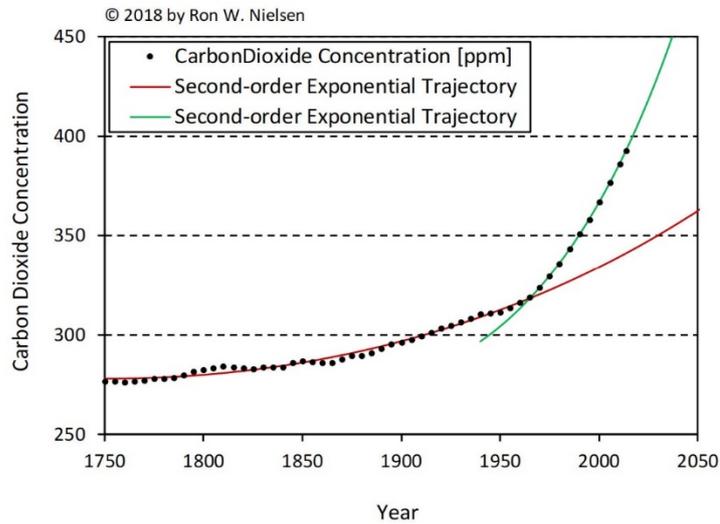

**Figure 22.** *Atmospheric concentration of carbon dioxide was initially following a slowly increasing and* accelerating *second-order exponential trajectory (* $a_0 = 1.481 \times 10^1$, $a_1 = -1.048 \times 10^{-2}$, $a_2 = 2.991 \times 10^{-6}$ *). From around 1965, it suddenly* accelerated *and started to follow a* faster *and* accelerating *second-order exponential trajectory (* $a_0 = 7.835 \times 10^1$, $a_1 = -7.721 \times 10^{-2}$, $a_2 = 2.049 \times 10^{-5}$ *). This sudden acceleration cannot be used in support of the concept of the Great Acceleration because atmospheric concentration of carbon dioxide is made of anthropogenic and natural components.*

## 3.13. Carbon dioxide emissions from burning fossil fuels

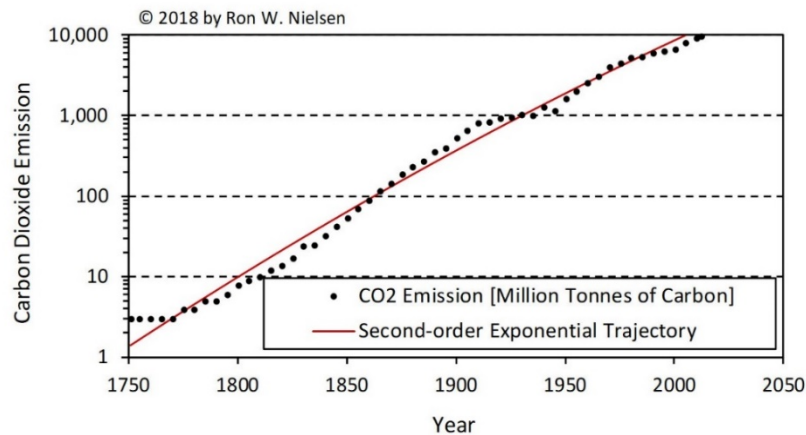

*Figure 23. Global emissions of carbon dioxide from burning fossil fuels (in million tonnes of carbon). Data were compiled by the Earth Policy Institute (EPI, 2013). From 1770, emissions of carbon dioxide were increasing but they were oscillating around a gradually* decelerating *second-order exponential trajectory (* $a_0 = -1.432 \times 10^2$, $a_1 = 1.230 \times 10^{-1}$, $a_2 = -2.346 \times 10^{-5}$ *). There was no Great Acceleration around 1950 or around any other time.*

## 3.14. Concentration of the atmospheric nitrous oxide (N$_2$O)

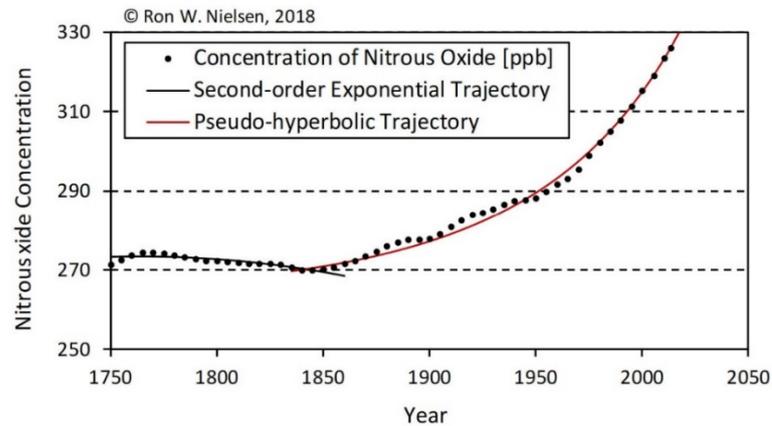

**Figure 24.** *Atmospheric concentration of nitrous oxide (N$_2$O), expressed in parts per billion [ppb], was initially decreasing along a second-order exponential trajectory ($a_0 = -4.727 \times 10^{-1}$, $a_1 = 6.898 \times 10^{-3}$, $a_2 = -1.956 \times 10^{-6}$). From around 1850, growth rate was increasing linearly with* the level of concentration, *generating a pseudo-hyperbolic growth ($C = -2.406 \times 10^{-14}$, $a_0 = -1.198 \times 10^{-2}$, $a_1 = 4.546 \times 10^{-5}$). Nitrous oxide concentration is made of natural and anthropogenic components. There was no sudden acceleration around 1950 but only around 1850.*

## 3.15. Concentration of the atmospheric methane (CH$_4$)

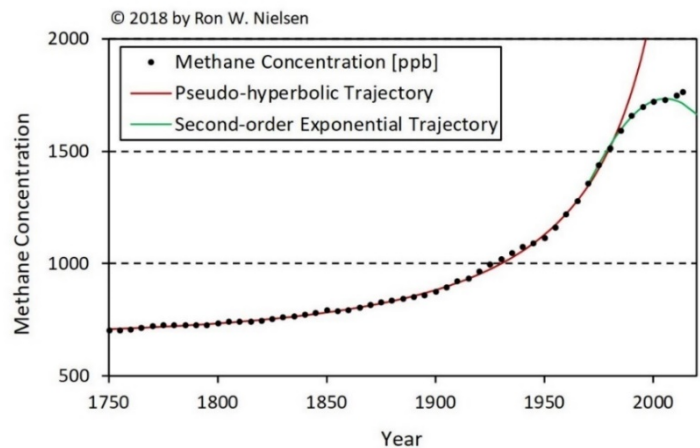

**Figure 25.** *Atmospheric concentration of methane (CH$_4$) in parts per billion [ppb]. Growth rate was initially increasing linearly with* the level of concentration *but from around 1980 it was decreasing linearly with time. Growth of methane concentration was initially pseudo-hyperbolic, described by the following parameters: $C = -5.007 \times 10^{-13}$, $a_0 = -1.072 \times 10^{-2}$ and $a_1 = 1.587 \times 10^{-5}$. However, from around 1980 it was following a* slower *and gradually* decelerating *second-order exponential trajectory ($a_0 = -7.863 \times 10^{2}$, $a_1 = 7.915 \times 10^{-1}$, $a_2 = -1.973 \times 10^{-4}$), reaching a predictable maximum in 2006. From 2006, it shows signs of a renewed increase. Methane emissions are made of natural and anthropogenic components. Nevertheless, there was no Great Acceleration around 1950 or around any other time.*

## *3.16. Loss of stratospheric ozone*

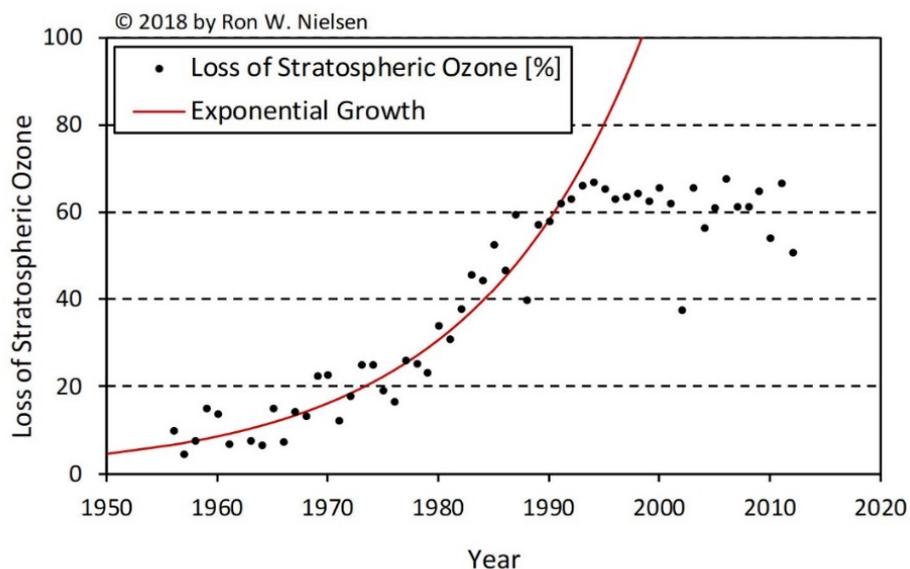

**Figure 26.** *Loss of stratospheric ozone was increasing exponentially ($c = 1.139 \times 10^{-54}$, $r = 6.446 \times 10^{-2}$) at the fast rate of 6.45% per year. From around 1992, gross rate is hard to analyse but it appears to be decreasing. Loss of stratospheric ozone appears to be steadily but slowly decreasing.*

## *3.17. Ocean acidification*

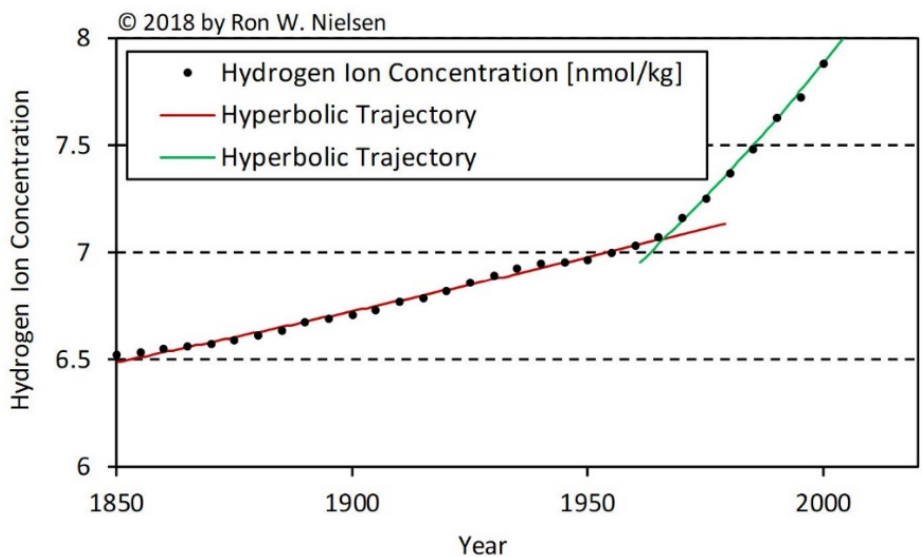

**Figure 27.** *Ocean acidification, described by the mean concentration of hydrogen ions ($H^+$) in nmol/kg, was increasing along a slow hyperbolic trajectory ($a = 3.553 \times 10^{-1}$, $k = 1.087 \times 10^{-4}$) but from around 1965 it suddenly* accelerated *and started to follow an approximately four times faster trajectory, as defined by the parameter $k$. Its parameters are: $a = 9.975 \times 10^{-1}$, $k = 4.353 \times 10^{-4}$. Ocean acidification is made of natural and anthropogenic contributions.*

## 3.18. Marine fish capture

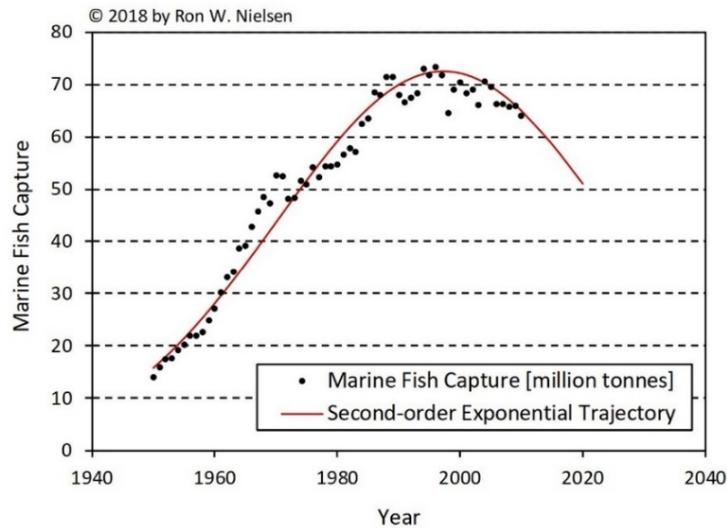

**Figure 28.** *Global marine fish capture (in million tonnes per year) was following a gradually decelerating second-order exponential trajectory ( $a_0 = -2.714 \times 10^3$, $a_1 = 2.722 \times 10^0$, $a_2 = -6.814 \times 10^{-4}$ ). It reached a predictable maximum in 1997 and stared to decline. There was never an abrupt acceleration but a continuing deceleration.*

## 3.19. Global shrimp production

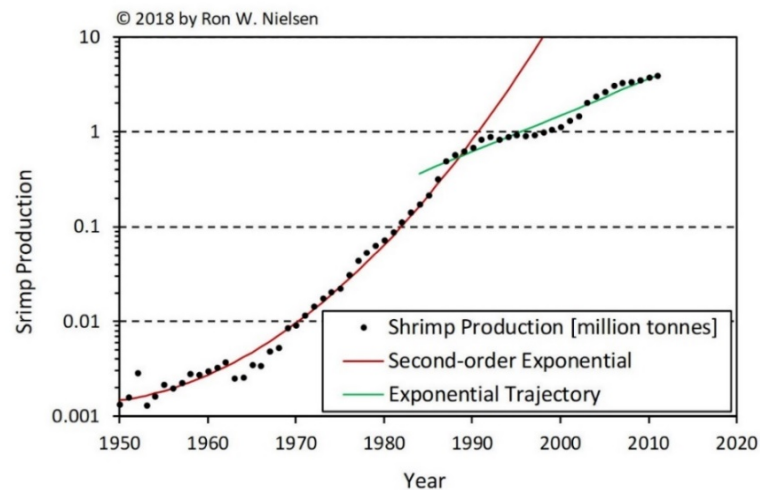

**Figure 29.** *Global shrimp production by aquaculture (in million tonnes). Production was increasing by following a continually accelerating second-order exponential trajectory ( $a_0 = 1.220 \times 10^4$, $a_1 = -1.255 \times 10^1$, $a_2 = 3.226 \times 10^{-3}$ ) until around 1990 when it decelerated and was diverted to a slower but continually accelerating exponential trajectory ( $c = 5.718 \times 10^{-77}$, $r = 8.798 \times 10^{-2}$ ). Its doubling time is 7.3 years. There was never a sudden acceleration but only a sudden deceleration in 1990 and a diversion to a slower but still accelerating trajectory.*

## 3.20. Global loss of tropical forests

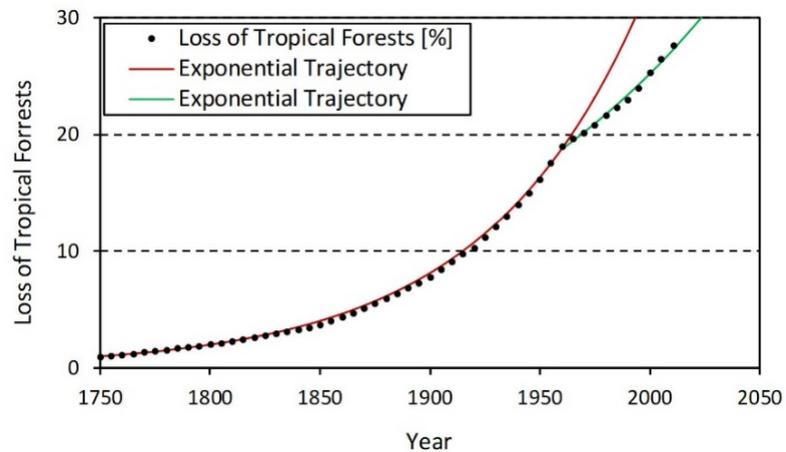

**Figure 30.** *Global loss of tropical forests (in percent of the forest area in 1700) was following a steadily accelerating exponential trajectory ($c = 2.337 \times 10^{-11}$, $r = 1.399 \times 10^{-2}$) until around 1960 when it was decelerated and diverted to a slower but still continually accelerating exponential trajectory ($c = 9.640 \times 10^{-6}$, $r = 7.388 \times 10^{-3}$). There was no Great Acceleration in 1950 or around any other time but there was a deceleration close to that year.*

## 3.21. Agricultural land area

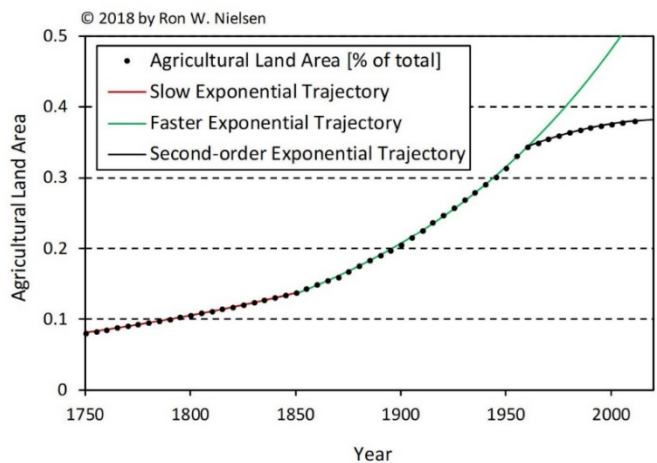

**Figure 31.** *Global agricultural land area (per cent of the total land area). Analysis of growth rate indicated three stages of growth: slow exponential to 1850 ($c = 7.357 \times 10^{-6}$, $r = 5.315 \times 10^{-3}$), faster exponential between 1850 and 1960 ($c = 2.301 \times 10^{-8}$, $r = 8.428 \times 10^{-3}$) and a slower, decelerating second-order exponential trajectory from 1960 ($a_0 = -1.007 \times 10^2$, $a_1 = 9.847 \times 10^{-2}$, $a_2 = -2.431 \times 10^{-5}$). The small acceleration from 0.5% per year to 0.8% per year, which occurred around 1850, was revealed only by studying growth rate. There was no Great Acceleration around 1950 or around any other time but only a small and hardly noticeable acceleration around 1850.*

## 4. Summary, discussion and conclusions

A summary of the mathematical analysis of anthropogenic signatures is presented in the Appendix. It includes not only signatures of purely anthropogenic origin but also signatures containing natural components. A briefer summary is presented in Table 2. It includes only anthropogenic signatures.

**Table 2**. *Search for the beginning of anthropogenic signatures*[a]

| Signature | Data From | Initial | Recent | Accel. | Decel. | Figures | Great Accel. |
|---|---|---|---|---|---|---|---|
| Population | AD 1 | A | D | 1950[b] | 1963 | 1, 3, 4 | X |
| GDP | AD 1 | A | A[c] | | 1950 | 2, 5 | X |
| GDP/cap | AD 1 | A | A[c] | | 1950 | 6, 7 | X |
| FDI | 1970 | A | U | | 2000 | 8, 9 | ND |
| Urban Pop. | 1750 | A | D | | 1960 | 10, 11 | X |
| Energy | 1750 | A | D | | 1950 | 12 | X |
| Fertilizers | 1900 | A | D | | 1972 | 13, 14 | X |
| Large Dams | 1750 | A | D | | 1965 | 15, 16 | X |
| Water | 1901 | A | A[c,d] | | 1979 | 17 | X |
| Paper | 1961 | D | D | | 1961 | 18 | ND |
| Transportation | 1963 | D | D | | 1963 | 19 | ND |
| Telecom. | 1960 | A | D | 1991[b] | 2000 | 20 | ND |
| Tourism | 1950 | D | D | | 1950 | 21 | ND |
| $CO_2$ from Fuels | 1751 | D | D | | 1770[g] | 23 | X |
| Ozone | 1950 | A | D | | 1992 | 26 | ND |
| Marine Fish | 1950 | D | D | | 1950 | 28 | ND |
| Shrimp Prod. | 1950 | A | A[c,e] | | 1990 | 29 | ND |
| Tropical Forests | 1750 | A | A[c,f] | | 1960 | 30 | X |
| Arable Land | 1750 | A | D | 1850[h] | 1960 | 31 | X |

Data From – data discussed in this publication; Initial – initial growth trajectory; Recent – recent growth trajectory; Accel. – sudden acceleration; Decel. – sudden and sustained deceleration; Great Accel. – Great Acceleration around 1950; FDI – Foreign Direct Investment; A – accelerating growth; D – decelerating growth; U – strongly unstable growth; X – no Great Acceleration around 1950; ND – no earlier data.

[a]) – Only anthropogenic signatures are listed. [b]) – Temporary boosting followed by a continuing deceleration. [c]) – Significantly slower than the initial trajectory. [d]) – Slow acceleration at the rate of only 1% per year. [e]) – Fast acceleration at the rate of 8.8% per year. [f]) – Slow acceleration at the rate of only 0.8% per year. [g]) – Emissions were constant between 1751 and 1770; [h]) – Exponential growth boosted from 0.5% to 0.8% per year but from 1960 there is a continuing deceleration.

### *4.1. The Great Deceleration*

A striking and *unexpected* result of the analysis presented here and summarised in Table 2, is the commonly occurring *decelerations* in the intensity of anthropogenic signatures in the second half of the 20th century. Analysis of data demonstrates consistently that *there was no Great Acceleration around 1950* but rather that there was a consistent, *en-masse deceleration* in growth trajectories, the phenomenon, which could be described as the *Great Deceleration*. Remarkably also, recent growth trajectories are generally *decelerating*, which gives hope for the future.

Recent anthropogenic signatures are characterised by a rapid increase, but a rapid increase should not be interpreted as a sudden acceleration without first checking that it was a sudden acceleration. Fast-increasing distributions, which initially increase slowly, can be misleading and it is easy to make mistakes with their interpretations. Indeed, it appears that they are often incorrectly interpreted. *Mistakes with their interpretations are made by even the most prominent and scrupulous scientists.* It is, therefore, essential to be always on guard when confronted with such distributions. They have to be always carefully and methodically analysed.

The abrupt Great Acceleration around 1950, or more generally in the second half of the 20th century did not happen, at least for the anthropogenic signatures listed in Table 2, i.e. for the signatures, which do not include the obvious natural contributions and which were published under the title of the Great Acceleration (IGBP, 2015). The sudden Great Acceleration is an illusion contradicted by data. In its place, mathematical analysis of anthropogenic signatures reveals the *Great Deceleration*. It was an unexpected but striking result.

It is remarkable that *all these anthropogenic signatures have been decelerated*. Most of them were diverted also to decelerating trajectories but some of them were diverted to accelerating trajectories. However, these new accelerating trajectories are not accelerating as rapidly as the original accelerating trajectories. There is now a greater chance that they will be also decelerated and diverted to decelerating trajectories. Notably, in this number of decelerated signatures there is the growth of global population and the global economic growth which, as suggested earlier, seems to reflect the *combined total* of all anthropogenic activities and impacts.

The term *Great Deceleration* is introduced here as a clear contrast with the term *the Great Acceleration*, used for precisely the same anthropogenic signatures (IGBP, 2015), to point out that there was no Great Acceleration but the Great Deceleration. Data demonstrate the *diametrically different interpretation* of growth trajectories than the interpretation based on impressions, showing that a correct interpretation of data can be obtained only by their rigorous analysis. This principle is well known in science but for some reasons it was not applied to the study of anthropogenic signatures and it resulted in their serious misinterpretation.

It is regrettable that with all the usual scientific precautions, the term *great acceleration* escaped scientific scrutiny, but there is always time to correct this mistake. Whatever reason might be produced to defend this term, its continuing use is not only undesirable but also scientifically unjustified. Even without specifying the time of the great acceleration, this term implies an acceleration at a certain time. It implies that the beginning of the Anthropocene can be determined by locating the beginning of this alleged great acceleration. It implies that anthropogenic signatures were increasing slowly in the past but their increase suddenly accelerated at a certain time and that it continues to accelerate, which is simply not true. *The perceived acceleration had no clearly marked beginning.* The so-called Great Acceleration had no clear beginning. Furthermore, this perceived acceleration was recently decelerated and diverted to slower trajectories. It is not just the growth of population but also consumption of natural resources that became recently diverted to slower and generally even to decelerating trajectories.

It is of concern that the concept of the Great Acceleration is not only accepted by scientific community but also presented to the general public without first checking it by a rigorous analysis of data. The article published on the Internet (IGBP, 2015) under the title *Great Acceleration* features prominently diagrams published earlier by Steffen et al. (2004) but now presented also under the title of *Great Acceleration*. Suitable hyperlinks are provided to assist in sharing this concept with a wide range of readers. This can be done as easily as by clicking a suitable link. It is an avalanche, which will be probably difficult to stop. There is also a natural human resistance against acknowledging mistakes and correcting them, a natural tendency to look for any possible excuse to avoid correcting a mistake.

When mistakes are made within scientific community, they can be easily corrected because scientists know that science is a self-correcting discipline, but if general public is involved, its members are less forgiving.

However, even within scientific community, promoting the concept of the Great Acceleration around 1950, or of a sudden acceleration at any other time, is undesirable. It would be embarrassing to submit a final proposal for the acceptance of the Anthropocene as a new geological epoch only to be told that the widely accepted concept of the Great Acceleration, promoted by the same group of people, is not based on science but on illusions reinforced by strongly desired anticipations. *If illusions can be so easily accepted, any other presented evidence can be also easily questioned.* "The greatest enemy of knowledge is not ignorance, it is the illusion of knowledge" – Stephen Hawking.

Concept of the Great Acceleration around 1950 is not based on a rigorous analysis of data but on impressions, but impressions can be misleading. "From these considerations then it is clear that the earth does not move, and that it does not lie elsewhere than at the centre" declared Aristotle in 350 BC (Aristotle, 2012, p. 14). However, what was so obviously and undeniably clear for Aristotle is now so obviously and undeniably incorrect.

Mistakes can be made in scientific investigations but mistakes are *always* corrected when they are positively identified. This is how science works. Sometimes it takes longer to correct them, but it is always preferable to correct them immediately.

The Great Acceleration understood as an abrupt and prominent increase in growth trajectories is an *illusion*. Data show that there was no such prominent and sustained abrupt increase. Great Acceleration interpreted in some other general sense is also an illusion because there were, and still are great decelerations. Emissions of carbon dioxide from fossil fuels, paper production, transportation, tourism and global marine fish capture have been following decelerating trajectories all the time, or at least for as long as indicated by available data. They were increasing but decelerating.

It is incorrect to interpret fast increase as an acceleration because fast increase could be also decelerating. It is incorrect to interpret anthropogenic signature as the Great Acceleration because some of them were decelerating all the time and all accelerating signatures have been decelerated and diverted to slower trajectories. The concept of the Great Acceleration is incorrect, imprecise and scientifically contradicted.

The term, *great acceleration*, is inaccurate and misleading. It describes an illusion, which does not explain anything and which contains incorrect information. The correct term is *the rapid increase*. It is not as dramatic as the great acceleration but describes accurately growth trajectories without suggesting the misleading, imprecise and incorrect interpretations.

The term *Great Acceleration* is impressive because it sounds so dramatic. However, its continuing use might turn out to be counterproductive when people discover that there was no great acceleration but great deceleration. The concept of the Anthropocene expresses the justified concern about human impacts on the environment but the concept of the Great Acceleration is an unnecessary aberration. Younger generations need to be warned but they also need to be encouraged. They need to have a correct, precise and reliable scientific information. They will have to be convinced that they can trust science. They also have to see that maybe not all is lost and that there is still hope for a sustainable future, and the evidence-based pattern, described here as the Great Deceleration, gives such a hope. There is still hope that humans will learn to live within their means. Rather than crash landing caused by the Great Acceleration, there might be a soft landing assisted by the Great Deceleration. This evidence-based interpretation of anthropogenic signatures could encourage various sections of community to increase their efforts in order to work harder towards a sustainable future.

## 4.2. The beginning

The search for the beginning of strong anthropogenic signatures produced negative results. Summary of results presented in Table 2 shows that even though the trajectories describing anthropogenic signatures are now, in general, decelerating, most of them were initially *accelerating*. Maybe we could even describe this almost common acceleration as a great acceleration but it would be an acceleration of a different kind than used in the concept of the Great Acceleration around 1950, which never happened. It is a great acceleration, which does not have a clearly marked beginning. What we see is a part of a great acceleration, which continued for thousands or even millions of years.

The Anthropocene is the age of humans but humans and their activities did not emerge suddenly in the past few hundred years but much earlier. Growth of human population was slow in the past but it was steadily increasing along predominantly steadily *accelerating* hyperbolic trajectories for at least 2,000,000 years (Nielsen, 2017c; see Figure 32). It was not a chaotic growth with many random ups and downs but a generally steady hyperbolic growth. Similar time dependence is also for the economic growth (De Long, 1998; Nielsen, 2017c). When growth of population and economic growth are very slow, as they were in the past, their ratio is approximately constant (2016i, 2017a; see also Figure 6) and economic growth is approximately directly proportional to the growth of population. The study of the dynamics of the growth of human population and of the economic growth shows that hyperbolic trajectories were exceptionally stable and robust.

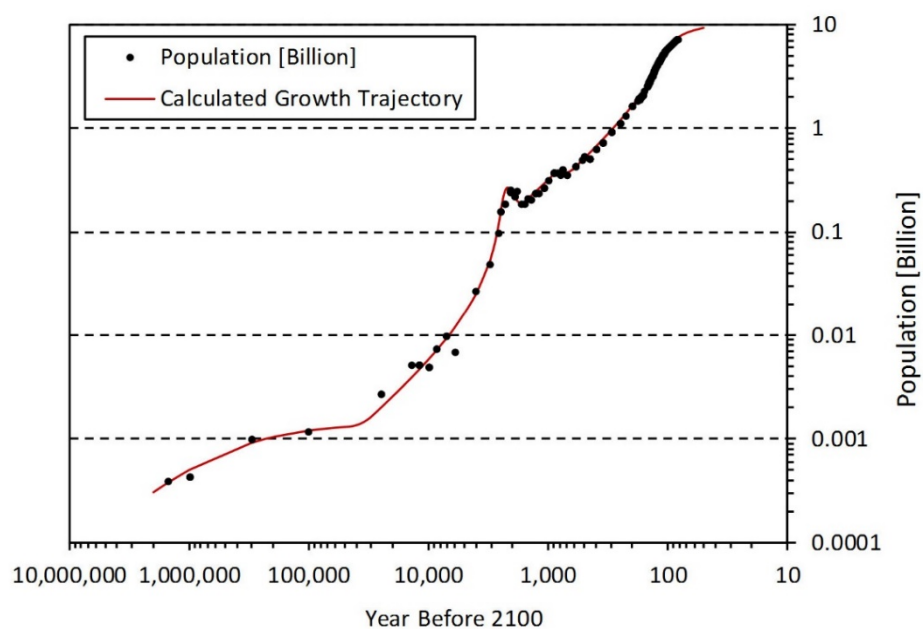

**Figure 32.** *Growth of human population in the past 2,000,000 years (Nielsen, 2017c)*

Hyperbolic growth is steadily accelerating and its acceleration is directly proportional to the size of the growing entity. For a long time, it was slow but it was gradually increasing until it reached the fast acceleration experienced in recent time. The recent fast acceleration in the growth of population and in the economic growth did not happen suddenly. Contrary to the commonly accepted interpretation of the growth of population, there was no sudden population explosion. There was also no sudden burst in the economic growth. The perceived sudden explosion is an illusion created by the strongly deceptive hyperbolic distributions (Nielsen, 2014, 2017f).

It is tempting to use the Industrial Revolution to mark the beginning of the Anthropocene. However, there are no suitable data, which could be used to carry out a detailed analysis of anthropogenic signatures before and after this event to check whether there was a sudden acceleration, but there are data for the key anthropogenic indicators: the growth of population and economic growth. They clearly show that Industrial Revolution had no impact on shaping their trajectories, even in Western Europe and even in the United Kingdom (Nielsen, 2014, 2016b, 2016e, 2016g, 2016l, 2017f; see also Figures 3-5). Industrial Revolution did not trigger a sudden acceleration in these two anthropogenic signatures. Consequently, *Industrial Revolution cannot be used to* determine *the beginning of the Anthropocene, but maybe it can be used to* define *it*.

The steady increase in the growth of human population and in the economic growth could be described as a great acceleration because it was not only increasing but also accelerating over a long time. It can be expected that the intensity of various anthropogenic activities and impacts was also increasing and maybe even accelerating, as suggested by a few recent examples shown in Table 2 and by the initially accelerating economic growth.

The steadily increasing growth of population and of economic growth in the past 2,000,000 years suggest that the Anthropocene evolved over a long time. It transcends the Pleistocene and Holocene epochs and it has no clearly marked beginning in recent years. Its beginning is lost in the mist of time, somewhere around 2,000,000 years ago or even earlier. The onset of hominine interaction with the environment could be perhaps traced to the emergence of the Oldowan tools around 2.5 million years ago, which was followed by countless other innovations, until their impacts became gradually strong and global, in much the same way as the growth of population and economic growth became also gradually so fast that they created an impression of a sudden explosion, but there was no sudden transition from slow to fast growth. How can we tell which of the gradually developing anthropogenic activities in the past contributed to the momentum of growth of the most recent trajectories? The present is strongly liked with the past and there is no clear demarcation line between the past and the present, except for the clear demarcation line determined by the recent *decelerations*.

The second half of the 20th century does not mark the beginning of the intensification of anthropogenic signatures, as suggested by the concept of the Great Acceleration, which is contradicted by data. On the contrary, the second half of the 20th century is marked by decelerations of the distributions describing the intensity of anthropogenic signatures. It is probably the *beginning of the end* of the strongest impacts and maybe even a gradual transition to a sustainable future.

Growth of population, which was steadily accelerating over the past 2,000,000 years has been decelerated and diverted to a slower trajectory from 1963. The steadily accelerating economic growth has been also decelerated and diverted to a slower trajectory from around 1950. It is, therefore, hardly surprising that the intensity of other anthropogenic signatures was also recently decelerated. However, as mentioned earlier, these results were unexpected. Reasons for the Great Deceleration could be investigated to understand its mechanism. It is also not known whether the observed Great Deceleration applies to other anthropogenic signatures, which were not listed in the IGBP sets of data. The ages-long process leading to the recent critically strong anthropogenic activities and impacts might be now coming to an end.

### *4.3. Is it a new geological epoch?*

"No one can deny the role of the human species in altering the global environment, but so did many biological innovations, like the first burrowing animals at the beginning of the Cambrian Period, and the rise of terrestrial forests in the Carboniferous" (Gehling, 2018). The problem is not with recognising the strong anthropogenic impacts on the environment but only in deciding whether recent human activities are causing a transition to a new geological epoch.

However, even if humans are now shaping a new geological epoch, it is still an ongoing process. It is probably too early to look for a proof of a transition from Holocene to Anthropocene. Maybe 1000 or 10,000 years from now, humans (if they survive) will look back and show that there was such a geological transition, but not now. Furthermore, the Great Deceleration seems to indicate that the power of shaping a new geological epoch by anthropogenic activities and impacts is now probably decreasing.

In order to have the Anthropocene recognised as a new geological epoch a search is now conducted for a convincing stratigraphic marker. However, this is precisely the problem with recognising the Anthropocene as a new geological epoch. There appears to be *no stratigraphic need* to use the concept of the Anthropocene as a new geological epoch. There is, however, a *need to find stratigraphic evidence* to support this concept. As noted by Finney and Edwards "the concept of the Anthropocene did not derive from the stratigraphic record" (Finney & Edwards, 2015, p. 6). There is no clear stratigraphic marker, which can be used to show convincingly a transition from Holocene to Anthropocene and it is probably even too early to look for such a marker. The way it is attempted to prove that Anthropocene is a new geological epoch is *first to define a marker* and then to look for it in stratigraphic deposits. This does not appear to be the way geological research is conducted.

There has been also too much attention focused on determining the beginning of the Anthropocene but so far, all attempts failed, and now we can understand why. The Anthropocene had no clear beginning in recent years and the widely accepted concept of marking this beginning by the beginning of the Great Acceleration is based on illusion because there was no Great Acceleration. There was no sudden acceleration in anthropogenic activities.

For the Anthropocene to be recognised as a new geological epoch, its beginning is far less important than a convincing stratigraphic evidence. It does not matter whether a transition from Holocene to Anthropocene occurred around the time of the Industrial Revolution, or in 1945, marking the first explosion of atomic bombs, or around 1950 the time of the alleged Great Acceleration that never happened, or at any other time during the second half of the 20th century. What matters is whether there was a transition. First, a convincing stratigraphic evidence has to be produced and only then one can worry about its timing. "Regrettably, focusing on the definition of the beginning of the Anthropocene can result in the lack of consideration of its stratigraphic content and its concept. It conveys the opinion that units of the geologic time scale are defined solely by their beginnings, rather than their content" (Finney & Edwards, 2015, p. 7).

If human activities are indeed shaping a new geological epoch, then transition to this new epoch is not marked by just one activity, such nuclear explosions or the production of plastics, but by *the whole range of anthropogenic activities*. A convincing stratigraphic marker caused by a single human activity will only show human presence but it will not prove that human presence caused a transition to a new geological epoch. If, for instance, a convincing stratigraphic evidence of nuclear explosions is found, it will only prove that there were nuclear explosions, which we already know, but it will not prove that nuclear explosions caused a transition from Holocene to Anthropocene. Perhaps humans are gradually causing a transition to a new geological epoch but this is still an ongoing process. Its geological impacts might perhaps become clear in a distant future, maybe through a collection of stratigraphic markers. It is probably too early to look for them now.

There is also a question of the time scale and time resolution. Geological changes occur over a long time. Attempts of finding a marker defining a transition from Holocene to Anthropocene appear to be based not only on the assumption that a single human activity, such as nuclear explosions or the production of plastics, could be used as an acceptable prove of such a geological transition but also that this transition occurred over only a few decades. Is it reasonable to expect that a transition to a new geological epoch would have occurred over such a short time or that it was caused by such a single and miniscule event? If the transition to a new geological epoch is ongoing, would it not be more reasonable to expect that a convincing evidence might be produced in the future?

Efforts of proving that the Anthropocene is a new geological epoch appear to be scientifically unjustified. The correct way of doing it is to wait for a clear and convincing stratigraphic evidence and only then to propose that there was a transition from Holocene to Anthropocene. This can happen only in the future.

However, is it really so vitally important to have the Anthropocene recognised as a new geological epoch? Nothing is going to change in the intensity of anthropogenic signatures if the Anthropocene is recognised as a new geological epoch. The Anthropocene might or might not be a new geological epoch. At present, it is just a convenient name to describe the recent strong anthropogenic impacts on the environment. It originated as a proposed name for a new geological epoch but so far, all attempts to prove that it is a new geological epoch were unsuccessful.

Analysis of the past growth of population and of the economic growth makes it clear why it is so formidably difficult to determine the beginning of the Anthropocene. There was simply no recent beginning. It is hard to do what is impossible to do. It is like trying to find a beginning of a circle. In the same category could be also the current efforts to prove that the Anthropocene is a new geological epoch. Maybe it is also impossible to do.

Research projects are not necessarily abandoned just because they are difficult. Difficult research projects can lead to important discoveries but the current attempts of proving that the Anthropocene is a new geological epoch does not appear to be promising.

### *4.4. Further research*

Two types of research activities related to the topic of the Anthropocene appear to be misdirected: (1) attempting to determine the beginning of the Anthropocene, because Anthropocene has no clear beginning in recent years, and (2) trying to prove that Anthropocene is a new geological epoch, because these efforts appear to be premature and scientifically unjustified.

The Anthropocene is real. For the first time in human history, humans have a profound *global* impact on the environment. For the first time, humans are shaping their own future and, to a certain degree, even the future of our planet. However, this profound global impact did not start at any particular time and it did not necessarily cause a transition to a new geological epoch. It evolved over a long time and it gradually became global. The Anthropocene can be recognised as a new phenomenon but its global impact has no clearly determinable beginning and it does not appear to mark a new geological epoch. It appears that the most important issue in studying the Anthropocene is to try to understand its mechanism.

The evidence supporting the concept of the Great Deceleration is overwhelmingly strong but it is limited. It might be possible that some other anthropogenic signatures, which are not listed in the IGBP publication, did not decelerate. Other data should be found and analysed. However, data for other signatures should be over a sufficiently long range of time to make a positive identification of a sudden acceleration or deceleration.

In contrast, the concept of the Great Acceleration around 1950 or in the second half of the 20th century is contradicted by a rigorous analysis of data. This concept originated also from an examination of a limited set of data but it was only a visual examination. It is a concept, which is based on impressions but now is shown to be contradicted by data, remarkably even by the same data, which were used in its support.

When looking for other examples of anthropogenic activities and impacts it is essential to understand that *fast-increasing trajectories should not be automatically interpreted as accelerating trajectories*. It is like driving a car fast and then trying to stop it. The car will be decelerating but the driving distance will be still increasing, and maybe even increasing fast.

Accelerating trajectories are characterised by a constant or increasing growth rate. The decelerating trajectories are characterised by a decreasing growth rate. If the growth rate is positive and decreasing, growth trajectory will be increasing. Depending on the value of the growth rate, it can be also increasing fast. Consequently, it would be incorrect to claim that every fast-increasing trend is accelerating. A trajectory might be rapidly increasing but not necessarily accelerating.

In order to see whether a given trend is accelerating or decelerating it is necessary to carry out mathematical analysis. Examples of fast-increasing anthropogenic signatures are meaningless unless it can be proven that they are represented by either accelerating or decelerating trajectories. Furthermore, as already mentioned, for a small range of data, it might be impossible to decide whether an accelerating trajectory is a result of an earlier deceleration. Table 2 lists some examples of such accelerating trajectories, which resulted from the earlier deceleration and economic growth is one of them. It was originally accelerating along a fast-increasing hyperbolic trajectory. It became decelerated and now is accelerating along a significantly slower trajectory, which is less critical than the earlier hyperbolic trajectory and which is easier to divert to a decelerating trajectory.

In discussions of other possible examples of anthropogenic signatures, distinction should be also made between desirable and undesirable trends. The accelerating undesirable trends reduce the probability of sustainable future but accelerating desirable trends increase its probability. Furthermore, even decelerating trends might be unsustainable. Trends can also change, to better or worse.

For instance, growth rate of the world population was steadily decreasing from 1963. The corresponding trajectory describing growth of population has been decelerating. The optimistically predicted maximum is around 12 billion (Nielsen, 2017e; United Nations, 2015). However, a more likely outcome is an asymptotic maximum of 15.6 billion (Nielsen, 2017e). Will any of these predicted maxima be sustainable? However, growth rate is now decreasing so slowly that it is almost constant. If it is going to remain constant, growth of the world population will be accelerating along an exponential trajectory and will be unsustainable.

Growth rate for the growth of population in China has been decreasing. Growth trajectory was decelerating but it was still increasing. However, from around 2008, growth rate started to hover around a constant value (World Bank, 2017). If this pattern is going to continue, growth of population in China will be accelerating along an exponential trajectory.

World economic growth has now settled around an accelerating exponential growth. However, its constant growth rate can, in due time, start to decrease when, for instance, economic stress is going to reach a high level. If this is going to happen, the current accelerating trajectory will be changed to a decelerating trajectory with a prospect of a sustainable future.

Another issue, which could be further investigated, apart from studying other anthropogenic signatures, is to try to explain the mechanism of the Great Deceleration. There could be various contributing factors such as improved environmental management, limits of Earth system and cross interaction between various anthropogenic activities and impacts. Each anthropogenic signature is also probably characterised by a different mechanism of growth. The distinction should be also made between potentially harmful and potentially beneficial human activities such as the increasing use of alternative sources of energy or the increasing use of electronic media, which gradually replaces paper production. All such studies could help to asses a chance for sustainable future.

Self-regulation is not necessarily imposed by the improvement in human interaction with Nature or by the limits and boundaries of the Earth system. Even with unlimited resources, there could be still a limit to growth. For instance, world economic growth was increasing along a hyperbolic trajectory but from 1950 it started to be diverted to a slower trajectory. This deceleration did not happen because all nations in the world agreed amiably and unanimously to stop following hyperbolic trajectory and to slow down their economic activities. There was no such mutual agreement. This deceleration was also not imposed by the critical boundaries of the Earth system because economic growth continues to increase and even to accelerate. The deceleration occurred spontaneously probably because it was simply impossible to cope with such a fast-hyperbolic growth. Economic growth started to follow a decelerating trajectory but gradually, its growth rate approached asymptotically a constant value, which describes exponential growth (Nielsen, 2015). It is a slower growth than the previous hyperbolic growth but accelerating. It is also an unsustainable growth and it will have to be terminated either by a diversion to a slower trajectory or by a collapse. Maybe this termination will be dictated by ecological limits but maybe not. When, in due time, exponential growth is going to lead to a high-intensity production and consumption, it will be no longer supported. There is a limit to how much can be produced and consumed over a certain time and this limit does not necessarily depend on the availability of natural resources or on a decision of some kind of an international economic tribunal. However, with limitations of natural resources, limit to growth can be reached earlier. Self-regulation might be helpful but controlled regulation is likely to produce better results.

It would be interesting, to study the mechanism of the Great Deceleration. What caused this remarkable transition. Was it a common cause or different causes for different signatures? It would be interesting to see whether similar decelerations are reflected in other anthropogenic signatures but most importantly it would be interesting to examine how the harmful and still accelerating trajectories could be also decelerated. It is a multidisciplinary challenge that could yield vital results.

One of the puzzling features revealed by the current analysis is the sudden acceleration in the atmospheric carbon dioxide concentration around 1965, which coincides with ocean acidification but is not correlated with the carbon dioxide emissions from burning fossil fuels. Reasons for this sudden acceleration are unclear.

Without a successful control of anthropogenic activities, there might be no Anthropocene. The Earth will survive without humans and so will also many other life forms. They will probably even thrive without humans. However, if humans are still around they will probably worry only about how to survive rather than about debating the beginning of the Anthropocene and proving that it is a new geological epoch.

## *4.5. Concluding remarks*

When the concept of the Anthropocene as a new geological epoch was introduced it was surprising and even disappointing to see that it was criticised. Evidence supporting this concept seemed to be overwhelmingly strong. For the first time in human history, human population was increasing so fast that it was even described as the population explosion. For the first time in human history, humans were changing not just local but global environment, not just changing it but changing it fast. All the signs were indicating that humans had an extraordinary impact on shaping not only their own future but perhaps also the future of this planet.

The concept of the Great Acceleration reinforced the concept of the Anthropocene as a new geological epoch but shifted the focus from the 1800s to the mid-1900s as the beginning of the explosive human impacts on the planet's environment. It was like having two explosions: one explosion around the 1800s and one around the 1950s, an apparent double confirmation that the Anthropocene could be a new geological epoch, except that the two explosions never happened, which illustrates how deceptive are the distributions associated with the concept of the Anthropocene. They have to carefully and methodically analysed; otherwise they lead easily to incorrect conclusions.

The misinterpretation of data was not caused by some kind of a well-coordinated conspiracy to prove that the Anthropocene is a new geological epoch with a well determinable beginning but by the genuine difficulty to understand these data. It is simply impossible to interpret them correctly without a mathematical analysis but mathematical analysis is also difficult. Only hyperbolic growth is trivially easy to analyse. Other distributions are not. However, even with hyperbolic growth one has to be careful to identify it correctly.

A brilliant scientist does not necessarily know how to analyse such data as describing anthropogenic signatures but a person who knows how to analyse them does not necessarily know how to interpret results of mathematical analysis. For instance, why was global water consumption diverted in around 1979 to such a dramatically slower trajectory? Why did global direct investments become so unstable from around 2000? Why was there a sudden increase in the atmospheric carbon dioxide concentration from around 1965 and why this sudden increase is not reflected in the carbon dioxide emissions from fossil fuels? Why was the atmospheric nitrous oxide concentration decreasing until around 1850? Why did the atmospheric methane concentration reach a maximum in around 2006 and why did it start to increase again? Why was the increase in the agricultural land area diverted in around 1960 to a significantly slower and decelerating trajectory?

With the concept of the Great Acceleration, demonstrated by a visual examination of data, it was easy to imagine that humans might have been causing a transition to a new geological epoch. If their past impact was strong, so strong that it was seen as a sudden explosion, it was just an introduction to an even greater explosion in the mid-1900s.

However, after studying the dynamics of the growth of population and of the economic growth it became clear that the interpretation of the Anthropocene has to be changed. This extensive study indicated that human impacts did not commence suddenly. The Anthropocene is a part of a longer human history. It has no recent beginning and it does not appear to be a part of the geological timeline. Furthermore, the intensity of human impacts is now moderated.

It is well known that growth of global population is now slowing down. Analysis of the economic growth has led to similar conclusions (Nielsen, 2016b, 2016e). Economic growth was also explosive, in much the same way as the growth of population, but both of them started to be diverted to slower trajectories, population from around 1963 and economic growth from around 1950. Not only the growth of population but also the total consumption of natural resources reflecting the combined intensity of anthropogenic signatures, started to be diverted to a slower trajectory.

Results of the analysis presented here confirmed the earlier results based on the study of the growth of population and of economic growth. A search of the Great Acceleration revealed, unexpectedly but convincingly, the presence of the Great Deceleration. In retrospect, these new results should have been expected. They are in harmony with the results of the earlier analysis of the growth of population and of economic growth. All these results suggest new interpretation of the Anthropocene. Not only was there no Great Acceleration but, in its place, there is a Great Deceleration. The momentum of anthropogenic activities and of their impacts is now decreasing. In general, distributions describing the intensity of various anthropogenic signatures are not yet decreasing but decelerating. Maybe the irreparable damage has been already done but maybe humans are gradually learning how to improve their interaction with Nature. The unexpected Great Deceleration, which needs to be further investigated, gives hope for the future. The decreasing momentum of anthropogenic impacts confirms and reinforces the reservations about the interpretation of the Anthropocene as a new geological epoch.

Mathematical analysis of data shows that the Anthropocene has no clearly marked beginning in recent years. It also puts in doubt whether the Anthropocene is a new geological epoch. However, these two features do not diminish the significance of the Anthropocene. The key issue is not in finding the beginning of the Anthropocene, which is impossible, or to prove that it is a new geological epoch, which is probably premature, but to understand this uniquely new phenomenon, to learn how to control its impacts, how to understand its mechanism and how to shape a sustainable future. It is a multidisciplinary challenge. Whether it had a recent beginning or not, whether it is a new geological epoch or not, the Anthropocene is here to stay and the question is whether it will continue to dictate human future or whether its impacts could be suitably moderated, maybe even controlled.

## Acknowledgements

I want to thank Cornelia Ludwig for making her compilation of data available on the website of the International Geosphere-Biosphere Programme. Without them, my analysis would have been much more difficult.

I also want to thank Will Steffen and his colleagues, for preparing their impressive and important set of diagrams and for publishing them in 2004. I want to thank Will Steffen for sharing these diagrams with me at the time when my book was published. Without these diagrams there would probably be no compilation of data illustrating anthropogenic signatures and no discovery of the Great Deceleration.

I want to thank Paul Crutzen for endorsing my book when after my retirement from research in nuclear physics I turned my attention to a study of critical events, which are shaping our future. I want to thank him for his long collaboration and for introducing me to the concept of the Anthropocene. It was a long journey, which in the past few years diverted me to a study of the dynamics of the growth of human population and of the economic growth, the two central processes shaping the Anthropocene.

I want to thank Jan Zalasiewicz for sharing with me his interesting article published in *Science*. It convinced me about the need to analyse the distributions describing the recent anthropogenic signatures. All these past encounters and events have gradually led to the presented here mathematical analysis.

I want to thank Matt Edgeworth for carefully reading my manuscript and for his most helpful constructive criticism. Using his suggestions, I was able to reduce the potential misinterpretations of my discussion.

I want to thank Dan deB. Richter and Jai Syvitski for their encouraging support, Jim Gehling and Michael Wagreich for their helpful comments and Stan Finney for confirming my reservations about claiming the Anthropocene as a new geological epoch.

It was particularly important to me to have a critique by Paul Crutzen, the Author of the concept of the Anthropocene. He has graciously described my study as a masterpiece and as a really impressive work.

This paper is for discussion. In order to preserve the scientific integrity of the concept of the Anthropocene it is essential to understand it correctly. This is a multidisciplinary field of research. If any of the arguments and explanations presented here are inaccurate, incorrect or questionable, they will be readily corrected. Anthropocene is an important topic that affects us all and it has to be correctly interpreted.

# Appendix

The following list contains a more detailed summary of the mathematical analysis of anthropogenic signatures.

1. Population:
    - Figures 1, 3 and 4
    - Data discussed her are from AD 1
    - Initially, hyperbolic growth (*accelerating*)
    - 1950 – minor temporary boosting
    - 1963 – growth *decelerated* and diverted to a *slower* and continually *decelerating* trajectory (Nielsen, 2017e; United Nations, 2015).
    - No Great Acceleration around 1950 or around any other time but only a minor temporary boosting followed by a continuing *deceleration*.
2. Gross Domestic Product (GDP):
    - Figures 2 and 5
    - Data discussed her are from AD 1
    - Initially, hyperbolic growth (*accelerating*)
    - 1950 – growth *decelerated* and diverted to a *slower* trajectory
    - The new trajectory was initially *decelerating* but is now approaching asymptotically an *accelerating* exponential growth.
    - No Great Acceleration around 1950 but a *deceleration* and a diversion to a slower growth, which is now again *accelerating*.
3. GDP/cap:
    - Figures 6 and 7
    - Data discussed her are from AD 1
    - Initially, linearly-modulated hyperbolic growth (*accelerating*)
    - 1950 – growth *decelerated* and diverted to a *slower* trajectory
    - No Great Acceleration around 1950 but *deceleration* around 1950 and diversion to a slower but still *accelerating* trajectory.
4. Foreign direct investment:
    - Data from 1970
    - Figures 8 and 9
    - Initially, second-order exponential growth (*accelerating*)
    - 2000 – diversion to a slower but strongly unstable pattern of growth
    - *No abrupt and sustained acceleration* at any time but a diversion to a *slower* pattern of growth.
5. Urban population:
    - Data from 1750
    - Figures 10 and 11
    - Initially, pseudo-hyperbolic growth (*accelerating*)
    - 1960 – growth *decelerated* and diverted to a slower and gradually *decelerating* trajectory
    - No Great Acceleration around 1950 or around any other time but a sustained *deceleration* from 1960.
6. Consumption of primary energy:
    - Data from 1750
    - Figure 12
    - Initially, growth hyperbolic (*accelerating*)

- 1950 – growth *decelerated* and diverted to a slower and gradually *decelerating* trajectory
- No Great Acceleration around 1950 or around any other time but a *deceleration* around that year and diversion to a continually *decelerating* trajectory.

7. Consumption of fertilizers
    - Data from 1900
    - Figure 13
    - Initially, second-order exponential growth (*accelerating*)
    - 1972 – growth *decelerated* and diverted to a *slower* trajectory
    - 1989 – growth reached unexpected maximum and started to decrease
    - 1993 – growth diverted to an even slower and continually *decelerating* second-order exponential trajectory
    - No Great Acceleration around 1950 but a *deceleration* around 1972.

8. Large dams:
    - Data from 1750
    - Figures 15 and 16
    - Initially, growth was increasing monotonically along a second-order exponential trajectory (*accelerating*)
    - 1965 – growth *decelerated* and diverted to a *slower* and continually *decelerating* logistic trajectory
    - No Great Acceleration around 1950 but *deceleration* around 1965 and diversion to a continually *decelerating* trajectory.

9. Water consumption:
    - Data from 1901
    - Figure 17
    - Initially, hyperbolic growth (*accelerating*)
    - 1979 - growth *decelerated* and diverted to a *slow* exponential trajectory (*accelerating* but very slowly at the rate of 1% per year)
    - *No Great Acceleration* around 1950 or around any other time but a *deceleration* in around 1979 and a diversion to a slower trajectory.

10. Paper production:
    - Data from 1961
    - Figure 18
    - From the beginning, second-order exponential growth continually *decelerating*
    - *No abrupt change* in the growth trajectory but a continuing *deceleration*.

11. Transportation:
    - Data from 1963
    - Figure 19
    - From the beginning, second-order exponential growth (*decelerating*)
    - *No major abrupt change* in the growth trajectory but a continuing *deceleration*.

12. Telecommunication:
    - Data from 1960
    - Figure 20
    - Initially, exponential growth until 1991 (*accelerating*)
    - 1991 – temporary boosting
    - 2000 – *deceleration* and a diversion to a continually *decelerating* trajectory tentatively described by a pseudo-logistic distribution, because growth in the number of landlines and subscriptions is not likely to decrease but more likely to approach an asymptotic maximum.

13. International tourism:
    - Data from 1950
    - Figure 21
    - From the beginning, second-order exponential growth (*decelerating*)
    - No abrupt acceleration in the growth trajectory but a continuing *deceleration.*
14. Atmospheric concentration of carbon dioxide:
    - Data from 1750
    - Figure 22
    - Anthropogenic and natural contributions
    - 1750-1965 – second-order exponential growth (*accelerating*)
    - From 1965 – abrupt *acceleration* and diversion to a faster second-order exponential trajectory (*accelerating*)
    - Acceleration around 1965 cannot be used in support of the concept of the Great Acceleration because carbon dioxide concentration contains natural contributions
15. Carbon dioxide emissions from burning fossil fuels:
    - Data from 1751
    - Figure 23
    - 1751-1770 – emissions constant
    - From 1770 – emissions *increasing* but along a continually *decelerating* second-order exponential trajectory
    - No Great Acceleration around 1950 but a continuing *deceleration* from around 1770.
16. Atmospheric concentration of nitrous oxide ($N_2O$):
    - Data from 1750
    - Figure 24
    - Anthropogenic and natural components
    - 1750-1850 – a decreasing, second-order exponential distribution (*decelerating*)
    - From 1850 – growth *accelerated* and diverted to a continually *accelerating* pseudo-hyperbolic trajectory
    - No sudden acceleration around 1950 but around 1850.
    - This distribution cannot be used in support of the Great Acceleration because it contains natural components.
17. Atmospheric concentration of methane ($CH_4$):
    - Data from 1750
    - Figure 25
    - Anthropogenic and natural contributions
    - Initially, pseudo-hyperbolic growth until 1990 (*accelerating*)
    - 1990 – *decelerated* and diverted to a slower continually *decelerating* trajectory
    - 2006 – growth reached a predictable maximum and now shows signs of renewed increase
    - No Great Acceleration around 1950
    - Future trajectory unpredictable because of the insufficient number of data
18. Loss of the stratospheric ozone:
    - Data from 1956
    - Figure 26
    - Initially, loss of stratospheric ozone was increasing exponentially, *accelerating* at the rate of 6.45% per year and doubling every 11 years.
    - From 1992 – slowly decreasing.
    - Future trajectory is hard to predict because of the combination of the poor-quality data and their short range.

19. Ocean acidification:
    - Data from 1850
    - Figure 27
    - Anthropogenic and natural contributions
    - 1850-1965 – slowly-increasing hyperbolic trajectory (*accelerating*)
    - From 1965 – a faster-increasing hyperbolic trajectory (*accelerating*)
    - No acceleration around 1950 but a clear *acceleration* around 1965, which cannot be used in support of the Great Acceleration because of the combination of natural and anthropogenic contributions to ocean acidification.
20. Marine fish capture:
    - Data from 1950
    - Figure 28
    - From the beginning, fish capture has been following a second-order exponential and continually *decelerating* trajectory
    - 1997 – predictable maximum followed by decline
    - There was no sudden acceleration at any time but a continuing *deceleration*.
21. Shrimp production:
    - Data from 1950
    - Figure 29
    - 1950-1990 – second-order exponential growth (*accelerating*)
    - 1990 – growth *decelerated* and diverted to a slower but still a fast-increasing exponential growth, *accelerating* at the high rate of 8.8% per year.
    - No sudden acceleration at any time to a faster trajectory but only a *deceleration* in around 1990 and a diversion to a slower but still accelerating trajectory.
22. Loss of tropical forests:
    - Data from 1750
    - Figure 30
    - From the beginning, loss of tropical forests was increasing exponentially, *accelerating* at the rate of 1.4% per annum
    - 1960 – loss *decelerated* and diverted to a *slower*, exponential trajectory, which is *accelerating* but at a slower rate of 0.7%
    - *No Great Acceleration* in 1950 or at any other time but *deceleration* in around 1960 to a slower but still *accelerating* trajectory.
23. Agricultural land area:
    - Data from 1750
    - Figure 31
    - 1750-1850 – slow exponential growth, *accelerating* at the rate of 0.5% per year
    - 1850 – growth *accelerated* to a slightly faster exponential trajectory characterized by the growth rate of 0.8% per year
    - 1960 – growth was *decelerated* and diverted to a slower and continually *decelerating* second-order exponential trajectory.
    - *No Great Acceleration* in 1950 but only *minor acceleration* around 1850, which can be detected only by the examination of the growth rate. This acceleration was followed in 1960 by *deceleration* to a slower and continually *decelerating* trajectory.